\documentclass[10pt,conference,letterpaper]{IEEEtran}

\usepackage{type1cm}     % type1 computer modern font
\usepackage{algorithm}     % algorithms
\usepackage{graphicx}     % advanced figures
\usepackage{subcaption}
\usepackage{epstopdf}
\usepackage{xspace}     % fix space in macros
\usepackage{balance}     % to better equalize the last page
\usepackage{booktabs}     % nicer tables
\usepackage[font=footnotesize,labelfont=sc]{caption}     % captions on top for tables
\usepackage[hyphens]{url}     % handle long urls
\usepackage[bookmarks, pdftex, colorlinks=false]{hyperref}     % clickable references
\usepackage[font=small, tableposition=bottom]{caption}     % captions on top for tables
\usepackage[section]{placeins}
\usepackage[show]{chato-notes}
\usepackage{balance}
\usepackage{amsmath,amssymb}
\usepackage[noend]{algpseudocode}
\usepackage{siunitx}
\usepackage{multirow}
\usepackage{mathtools}
\DeclareGraphicsExtensions{.pdf,.png,.jpg,.eps}
\graphicspath{{./img/}}
\algdef{SE}[DOWHILE]{Do}{doWhile}{\algorithmicdo}[1]{\algorithmicwhile\ #1}%

\newcommand{\spara}[1]{\smallskip\noindent{\bf #1}}

\newenvironment {squishlist}
{\begin{list}{$\bullet$}
  { \setlength{\itemsep}{0pt}
     \setlength{\parsep}{3pt}
     \setlength{\topsep}{3pt}
     \setlength{\partopsep}{0pt}
     \setlength{\leftmargin}{1.5em}
     \setlength{\labelwidth}{1em}
     \setlength{\labelsep}{0.5em} } }
{\end{list}}

\def\baselinestretch{0.895}
\newcommand{\load}[2]{\ensuremath{L_{#1}(#2)}\xspace}
\newcommand{\staticload}[1]{\ensuremath{L_{#1}}\xspace}

\newcommand{\tuple}[1]{\ensuremath{\langle #1 \rangle}\xspace}
\newcommand{\hash}[1]{\ensuremath{\mathcal{H}_{#1}}\xspace}
\newcommand{\hashk}{\ensuremath{\mathcal{H}}\xspace}
\newcommand{\hashf}[1]{\ensuremath{\mathcal{H}(#1)}\xspace}

\newcommand{\utilization}[2]{\ensuremath{\mathcal{U}_{#1}(#2)}\xspace}

\newcommand{\workerqueue}[1]{\ensuremath{\mathcal{Q}_{#1}}\xspace}
\newcommand{\pei}{\textsc{pei}\xspace}
\newcommand{\peis}{{\pei}s\xspace}
\newcommand{\pe}{\textsc{pe}\xspace}
\newcommand{\pes}{{\pe}s\xspace}
\newcommand{\potc}{\textsc{p\textup{o}tc}\xspace}
\newcommand{\porc}{\textsc{p\textup{o}rc}\xspace}
\newcommand{\ch}{\textsc{ch}\xspace}

\newcommand{\cg}{\textsc{cg}\xspace}
\newcommand{\dspe}{\textsc{dspe}\xspace}
\newcommand{\dspes}{{\dspe}s\xspace}
\newcommand{\dagr}{\textsc{dag}\xspace}

\newcommand{\dagrs}{\textsc{dag}s\xspace}

\newcommand{\pkgs}{\textsc{pkg}\xspace}
\newcommand{\kg}{\textsc{kg}\xspace}
\newcommand{\sg}{\textsc{sg}\xspace}

\newcommand{\sources}{\ensuremath{\mathcal{S}}\xspace}
\newcommand{\numsources}{\ensuremath{s}\xspace}
\newcommand{\workers}{\ensuremath{\mathcal{W}}\xspace}
\newcommand{\worker}{\ensuremath{w}\xspace}
\newcommand{\colors}{\ensuremath{c}\xspace}
\newcommand{\numworkers}{\ensuremath{n}\xspace}
\newcommand{\key}{\ensuremath{j}\xspace}
\newcommand{\keyspace}{\ensuremath{\mathcal{K}}\xspace}
\newcommand{\capacity}[1]{\ensuremath{{c}_{#1}}\xspace}

\newcommand{\finishtime}[1]{\ensuremath{{\phi}_{#1}}\xspace}
\newcommand{\partitioner}{\ensuremath{{\mathcal{H}}}\xspace}
\newcommand{\mycomment}[1]{}
\newcommand{\real}{\ensuremath{\mathbb{R}}\xspace}
\newcommand{\keysize}{m\xspace}
\newcommand{\messageidentifier}{\ensuremath{i}\xspace}

\DeclareRobustCommand{\calD}[0]{{\mathcal D}}

\DeclareRobustCommand{\calC}[0]{{\mathcal C}}
\DeclareRobustCommand{\calK}[0]{{\mathcal K}}
\newcommand{\code}[1]{{\textsc #1}}

\newcommand{\BigO}[1]{\ensuremath{\operatorname{O}\left(#1\right)}}

\DeclareRobustCommand{\calD}[0]{{\mathcal D}}

\DeclareMathOperator*{\expect}{\mathbb{E}}
\DeclareMathOperator*{\avg}{avg}
\DeclareMathOperator*{\argmin}{argmin}
\DeclareMathOperator*{\argmax}{\arg\max}

\newtheorem{theorem}{Theorem}[section]

\newtheorem{lemma}[theorem]{Lemma}

\newtheorem{problem}[theorem]{Problem}

\DeclarePairedDelimiter\ceil{\lceil}{\rceil}

\usepackage[square,numbers]{natbib}     % better references
\title{Load Balancing for Skewed Streams on Heterogeneous Clusters}
\author{
\IEEEauthorblockN {
Muhammad Anis Uddin Nasir{\small $^{\$\#}$},
Hiroshi Horii{\small $^{\$}$},
Marco Serafini{\small $^{*}$},
Nicolas Kourtellis{\small $^{\ddagger}$} \\
Rudy Raymond{\small $^{\$}$},
Sarunas Girdzijauskas{\small $^{\#}$}, 
Takayuki Osogami{\small $^{\$}$}
}
\fontsize{10}{10}\selectfont\itshape
\small{$^{\#}$}Royal Institute of Technology, Sweden \hspace{2mm}
\small{$^{\$}$}IBM Research Tokyo, Japan \hspace{2mm}
\small{$^{*}$}Qatar Computing Research Institute \hspace{2mm}
$^{\ddagger}$Telefonica Research \\
 \fontsize{9}{9}\selectfont\ttfamily\upshape
 \vspace{-0.5mm}
anisu@kth.se,
horii@jp.ibm.com,
mserafini@qf.org.qa,
nicolas.kourtellis@telefonica.com \\
rudyhar@jp.ibm.com, 
sarunasg@kth.se,
osogami@jp.ibm.com
}

\begin{document}
\maketitle

%% ABSTRACT
\begin{abstract}

Streaming applications frequently encounter skewed workloads and execute on heterogeneous clusters.
%Primitive partitioning strategies for streaming applications operate efficiently under the assumptions: the resources are homogeneous and the messages are drawn from a uniform key distribution. % and the service time for the tuples follows a uniform distribution.
%These assumptions are often not true for the real-world use cases.
%Such adverse conditions require inferring the resource capacities and input distribution at run time. 
%to produce a fair assignment to tuples to the workers.
Optimal resource utilization in such adverse conditions becomes a challenge, as it requires inferring the resource capacities and input distribution at run time.
%However, gathering these statistics and finding an optimal placement often become a challenge when microsecond latency is desired.
% at the sources and requires probing the workers periodically to maintain the statistics.
In this paper, we tackle the aforementioned challenges by modeling them as a load balancing problem.
%In this paper, we address the load balancing problem for streaming engines running on a heterogeneous cluster and processing skewed workload. 
We propose a novel partitioning strategy called \emph{Consistent Grouping} (\cg), which enables each processing element instance (\pei) to process the workload according to its capacity.
%In constrast to the existing approaches that assign the stream to \peis randomly, \cg achieves near-optimal resource utilization by considering both workload and capacity distribution.
The main idea behind \cg is the notion of small, equal-sized virtual workers at the sources, which are assigned to workers based on their capacities. %  is to deal with load imbalance on heterogeneous resources and workloads with load-aware (local) bottom up approach, rather than not load-oblivious or top-down (global) approaches.
%The main idea behind \cg is to divide the workload into small and equal-sized bins, which are assigned to workers based on their capacity.
%that takes a radically different approach and delegates the load balancing problem to the processing element instances (\peis).
%This simple strategy transparently reduces the distributed problem to a local one, and enables each \pei to process the workload according to its capacity.
%The main idea is to allow idle \peis to steal the work from the overloaded \peis by negotiating the workload based on their available capacity.
We provide a theoretical analysis of the proposed algorithm and show via extensive empirical evaluation that our proposed scheme outperforms the state-of-the-art approaches, like key grouping. % with several orders of magnitude. % showing its effectiveness and applicability.
In particular, CG achieves 3.44x better performance in terms of latency compared to key grouping. %stateful grouping strategy.
%In particular, \cg achieves $3.44$x superior performance in terms of latency compared to state-of-the-art solutions. %key grouping, which is the state-of-the-art grouping strategy for stateful streaming applications. 
%Compared to key grouping, it improves the throughput of an example application on real-world datasets by up to $2$x, reduces the latency by $3.5$x
%\notes{Can we give any numbers? Similar to the ones we will report in the end of Introduction}
%\notes{which state-of-art? Can we be specific?}
\end{abstract}

%% INTRO
\section{Introduction}
\label{sec: intro}
%Streaming Model
Distributed stream processing engines (\dspes) have recently gained much attention due to their ability to process huge volumes of data with very low latency on clusters of commodity hardware.
\dspes enable processing information that is produced at a very fast rate in a variety of contexts, such as IoT applications, software logs, and social networks.
For example, Twitter users generate more than \num{380} million tweets per day\footnote{\url{http://www.internetlivestats.com/twitter-statistics/}} and Facebook users upload more than \num{300} million photos per day\footnote{\url{http://www.businessinsider.com/facebook-350-million-photos-each-day-2013-9}}. %has over 1.15 billion mobile daily active users .
%\notes{any references, even web-articles reporting these?}

Streaming applications are represented by directed acyclic graphs (\dagrs), where vertices are called \emph{processing elements} (\pes) and represent operators, and edges are called \emph{streams} and represent the data flowing from one \pe to the next.
For scalability, streams are partitioned into sub-streams and processed in parallel on replicas of \pes called \emph{processing element instances} (\pei).

%Applications
%Streaming applications perform light-weight operations such as filtering, aggregating, or joining, on the incoming data streams, to analyze the information in real time.
%For example, a typical application is the detection of trending hashtags in a stream of tweets.
%In this case, the \peis responsible for counting the occurrences of the hashtags trending hashtags also receive a predominant share of the messages in the stream.
%%As another example, when computing the reach of an information cascade in a social network, the \peis responsible for processing the few highly viral instances will follow a similar pattern.
%%%As another example, when counting the number of friends of users in a social graph, the \peis responsible for processing the few high-degree nodes will observe the same pattern.
%The same behavior can be observed in other domains such as classification (by grouping on classes and attributes), statistical language models (grouping on words), and streaming graph processing (grouping on vertices).

\begin{figure}[t]
\begin{center}
\includegraphics[width=0.75\columnwidth]{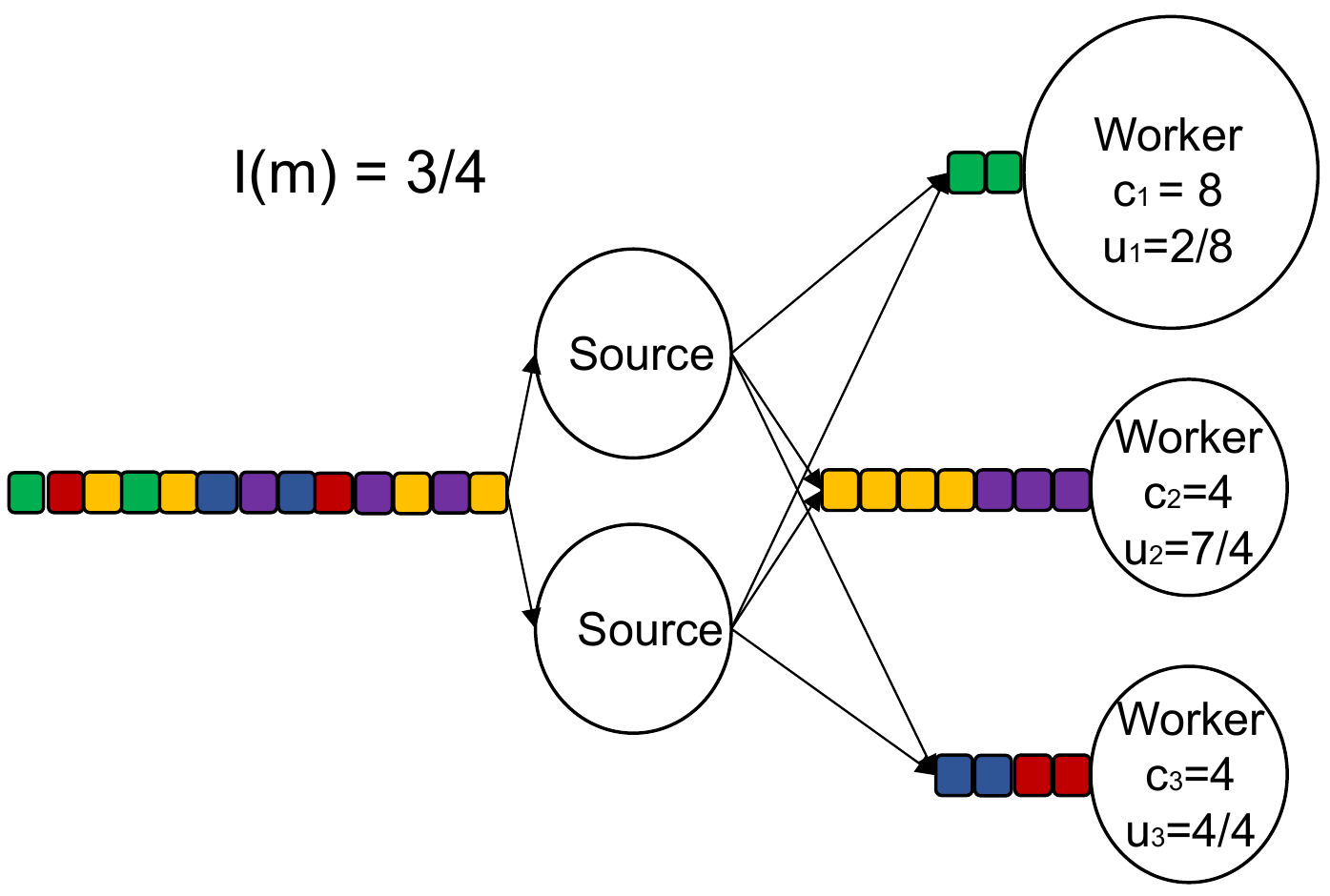}
\caption{Example showing that key grouping generates imbalance in the presence of a heterogeneous cluster.
The capacity and the resource utilization of the $i$-th worker is represented by $c_i$ and $u_i$ respectively. 
Each key ($\key\in \keyspace$) is represented with different color box.
Imbalance $I(m)$ is the difference between the maximum and the average resource utilization (see section 
\ref{sec:preliminaries} for details).}
\label{fig:kg-example}
\end{center}
\vspace{-\baselineskip}
\end{figure}

%current state -> hashing
%\spara{Challenges.}
Applications of \dspes, especially in data mining and machine learning, typically require accumulating state across the stream by grouping the data on common fields \cite{ben-haim2010spdt,berinde2010heavyhitters}.
Akin to MapReduce, this grouping in \dspes is usually called \emph{key grouping} (\kg) and is implemented using hashing \cite{nasir2015power}.
\kg allows each source \pei to route each message solely via its key,  without needing to keep any state or to coordinate among \peis.
However, \kg is unaware of the underlying skewness in the input streams \cite{lin2009curse}, which causes a few \peis to sustain a significantly higher load than others, as demonstrated in Figure \ref{fig:kg-example} with a toy example.
This sub-optimal load balancing leads to poor resource utilization and inefficiency.

%Heterogenous clusters
The problem is further complicated when the underlying resources are heterogeneous \cite{koliousis2016saber, schneider2016dynamic} or changing over time \cite{zaharia2008improving, suresh2015c3}.
For various commercial enterprises, the resources available for stream mining consist of dedicated machines, private clouds, bare metal, virtualized data centers and commodity hardware. 
For streaming applications, the heterogeneity is often invisible to the upstream \peis and requires inferring the resource capacities in order to generate a fair assignment of the tasks to the downstream \peis. %, as shown in Figure \ref{fig:kg-example}.
%Such information requires probing the downstream \peis periodically to maintain the statistics.
However, gathering statistics and finding optimal placement often leads to bottlenecks, while at the same time microsecond latencies are desired \cite{kalyvianaki2016themis}. 

%round robin
Alternatively, stateless streaming applications, like interaction with external data sources, employ \emph{shuffle grouping} (\sg) to break down the stream load equally to each of the \peis, i.e., by sending a message to a new \pei in cyclic order, irrespective of its key.
\sg allows each source \pei to send equal number of messages to each downstream \pei,  without the need to keep any state or to coordinate among \peis.
However, similarly to \kg, \sg is unaware of the heterogeneity in the cluster, which can cause some \peis to sustain unpredictably higher load than others.
Further, \sg typically requires more memory to express stateful computations \cite{nasir2015power,katsipoulakis2017holistic}.

%\spara{Problem Statement.}
In this present work, we study the load balancing problem for a streaming engine running on a heterogeneous cluster and processing non-uniform workload.
To the best of our knowledge, we are the first to address both challenges together.
We envision a light-weight and fair key grouping strategy for both stateless and stateful  streaming applications. %\notes{what set? Ideally you have talked about them earlier and now you say which types of applications you tackle (e.g., stateful, stateless, etc.)} %for \dspe running on a cluster with heterogeneous machines. %based on their capacity.
Moreover, this strategy must limit the number of workers processing each key, which is analogous to reducing the memory footprint and aggregation cost for the stateful computation \cite{nasir2015power,katsipoulakis2017holistic}.
Towards this goal, we propose a novel grouping strategy called \emph{Consistent Grouping} (\cg), which handles both the potential skewness in input data distribution, as well as the heterogeneity in resources in \dspes.
%The main idea behind \cg is to divide the workload into small and equal-sized bins\footnote{We use the bins and virtual workers interchangeably throughout this paper to refer to virtual workers.}, which are assigned to workers based on their capacity.
\cg borrows the concept of virtual workers from the traditional consistent hashing \cite{godfrey2004load,godfrey2005heterogeneity} and employs rebalancing to achieve fair assignment, similar to \citep{shah2003flux,gedik2014partitioning,balkesen2013adaptive,castro2013integrating, suresh2015c3}.
In summary, our work makes the following contributions:
\begin{squishlist}
%\item We address the load balancing problem for \dspes running on heterogeneous cluster and processing skewed workload.
\item We propose a novel grouping scheme called Consistent Grouping to improve the scalability for \dspes running on heterogeneous clusters and processing skewed workload. %that provides fair assignment of messages to downstream operators.% by adapting to traditional consistent hashing.
\item We provide a theoretical analysis of the proposed scheme and show the effectiveness of the proposed scheme via extensive empirical evaluation on synthetic and real-world datasets. % of the proposed grouping to show its effectiveness.
In particular, \cg achieves bounded imbalance and generates almost optimal memory footprint. % close to optimal solution.
\item We measure the impact of \cg on a real deployment on Apache Storm. Compared to key grouping, it improves the throughput of an example application on real-world datasets by up to $2$x, reduces the latency by $3.44$x. % and avoid message drops. %failures \notes{failures? where did this come from?}.
\end{squishlist}

%% INTRO
\section{Overview of the Approach}
\label{sec: overview }
Consistent grouping relies on the concept of virtual workers and allows variable number of  \emph{virtual workers} for each \pei.
The main idea behind \cg is to assign the input stream to the virtual workers in a way that each virtual worker receives approximately the same number of messages.
Later, these virtual workers are assigned to the actual workers based on their capacity.
We refer to downstream \peis as workers and to upstream \peis as sources throughout the paper.
Similar approaches have been considered in the past in the context of distributed hash tables \cite{godfrey2004load, godfrey2005heterogeneity}.
\cg allows an assignment of tasks to \peis based on the capacity of the \peis.
Thus, the powerful \peis are assigned more work compared to less powerful \peis.
Next, we provide an overview of  \cg's components.% that allow its adaptation in the streaming context.
%The adaptation of \cg in the streaming context requires answering several challenging questions:
%1) How do we divide the messages into equal-sized virtual workers?
%, 2) How do we identify the imbalance in load on workers due to heterogeneity of their resources?
%and 3) How do we plan the migration? %, and (4) How to ensure that the system operates in a stable manner.

\begin{figure}[t]
\begin{center}
\includegraphics[width=0.85\columnwidth]{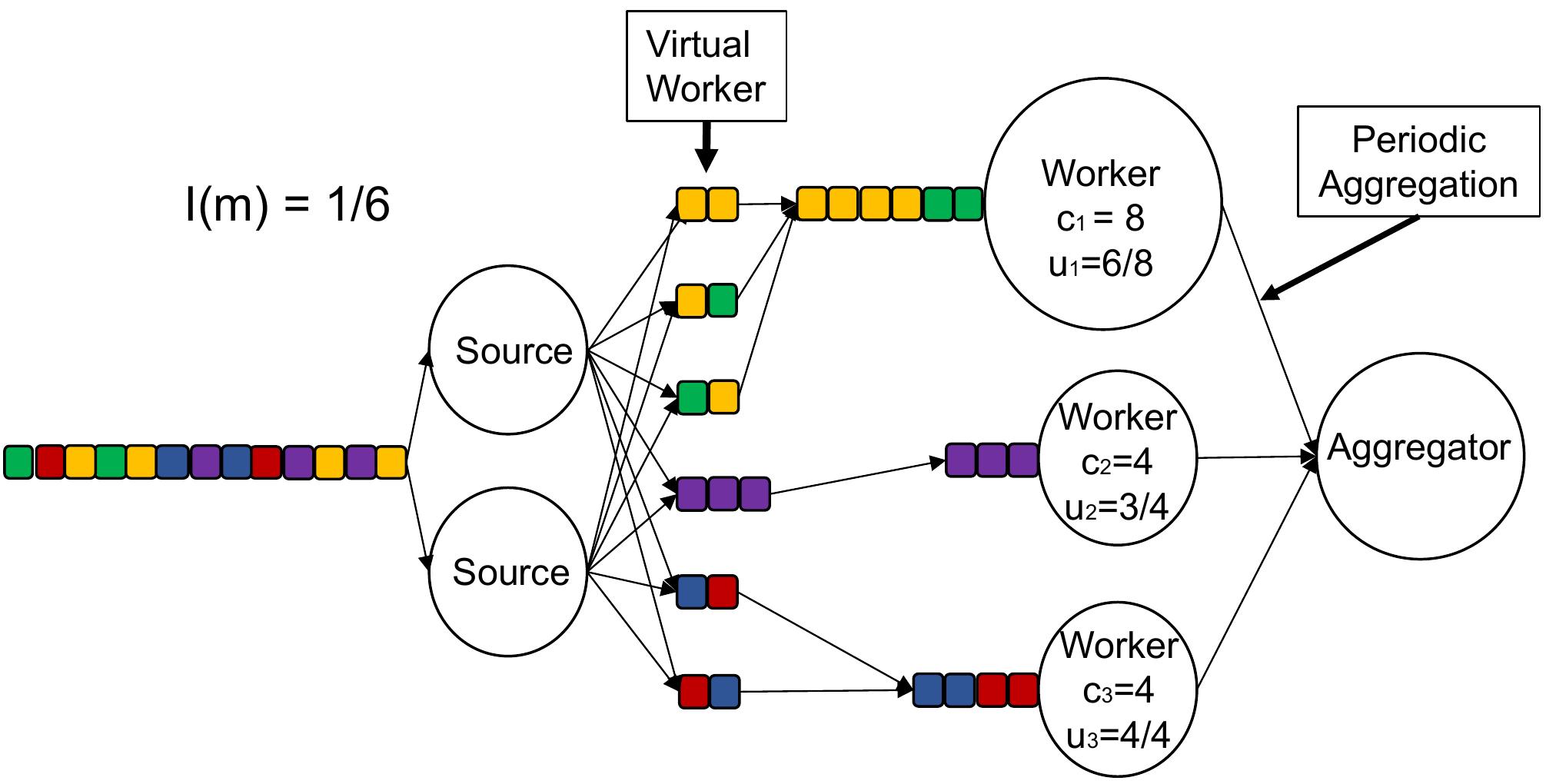}
\caption{Example showing that consistent grouping improves the imbalance in the presence of heterogeneous cluster, compared to key grouping.
The capacity and the resource utilization of the $i$-th worker is represented by $c_i$ and $u_i$ respectively. 
Also, each key ($\key\in \keyspace$) is represented with different color box. % and the width of a box represents the service time of the key.
Imbalance $I(m)$ is the difference between the maximum and the average resource utilization.}
\label{fig:cg-example}
\end{center}
\vspace{-\baselineskip}
\end{figure}

First, we propose a novel strategy called \emph{power of random choices} (\porc), which assigns the incoming messages to a set of equal sized virtual workers.
%\notes{is the size of the set fixed?}
The basic idea behind this scheme is to introduce the notion of capacity for the virtual workers. 
In particular, we set the capacity of each virtual worker to the average load $\times$ (1+$\epsilon$), for some parameter $\epsilon$.
Note that the capacity is calculated at run time using the average load.
Given a sequence of virtual workers for each key, \porc maps a key to the first virtual worker with a spare capacity.
%Given infinitely many hash functions that produce fixed set of choices for a assignment of a message to a virtual worker, \porc maps a key to the first virtual worker with a spare capacity.
%The candidate workers are produced using the hash functions.
\porc allows the heavy keys to spread across the other virtual workers, thus reducing the memory footprint and the aggregation cost. % compared to \potc.
The $\epsilon$ parameter in the algorithm provides the trade off between the imbalance and memory footprint.

Second, \cg takes a radically new approach towards load balancing and allows \peis to decide their workload based on their capacities.
We call this component as \emph{worker delegation}.
Each worker monitors its workload and sends a binary signal (increase or decrease workload) to the sources in case it experiences excessive workload.
This simple modification changes the distributed load balancing problem to a local decision problem, where each \pei can choose its share of workload based on its current capacity.
Moreover, worker delegation provides the flexibility to implement various application-specific requirements at each \pei.
%The upstream \peis receive two types of signals: a) increase workload and b) decrease workload. 
The sources react to the signals by moving virtual workers from one \pei to another. %\textbf{ins} the \peis. % swap virtual workers and shrink a virtual worker. %spawning and killing the virtual \peis.
Note that it is required that sources receive the signal and operate in a consistent manner, performing the same routing of messages.
%The process of instantiating additional virtual \peis is equivalent to increasing the share of future workload. % keys in the future.
%Similarly, killing existing virtual \peis allows overloaded \peis to reduce the amount of assigned workload. 
%Each downstream \pei requires updating the upstream \peis in case of experiencing undesirable workload.
Such an operation might negatively impact the performance of a streaming application, as it requires one-to-many (from one worker to all sources) broadcast messages across the network.
To overcome this challenge, we relax the consistency constraint in the \dagr and allow sources to be eventually consistent.
Specifically, we propose \emph{piggybacking} that allows encoding the binary signals along with the acknowledgment message to avoid extra communication overhead. %, we propose to \emph{encode} the binary signals from the downstream operators along with the acknowledgement of the messages. 
%During the execution, the upstream operators only receive the signal from the downstream operator as a response to messages.
%This means that few \pei might continue receiving the messages with the same key even after taking the decision to offload few keys. 
%However, in worst case the number of messages received by downstream \peis is upper bounded by the number of upstream operators, assuming no network delays. 

%Lastly, we need to take care of the message migration, as a message with the same key might be forwarded to different \peis.
Lastly, \cg ensures that each message is processed in a consistent manner by discarding the message migration phase. 
When a source receives a request to change (increase or decrease) the workload, \cg relocates virtual workers assigned to the overloaded worker, thus, only affecting the future routing of the messages.
%Concretely, each worker processes the messages that are assigned to it; changes in the routing only affect the messages that arrive in a later time.
\cg follows the same programming primitive as Partial Key Grouping (\pkgs) \cite{nasir2015power} for stream partitioning, supporting both stateless and stateful map-reduce like applications. % and not every algorithm can be expressed with it.
%As stateless application do not require any state migration, supporting stateless application is trivial.
%Stateful applications typically require accumulating state across the stream by grouping the data on common fields \cite{ben-haim2010spdt,berinde2010heavyhitters}.
We propose \emph{periodic aggregation} to support map-reduce like stateful applications, which leverages the existing \dagr and imposes a very low-overhead in the stream application.
%Clearly, consistent grouping is capable of generating fair assignment compared to key grouping. % while maintaining the semantics of single state per key.
%Note that the distribution shown in the figure corresponds to the assignment after a single time bucket, which allows the sources to change the assignment after the first time bucket.
Figure \ref{fig:cg-example} provides an example using \cg for the \dagr in Figure \ref{fig:kg-example}.
%Consistent grouping enables fair assignment of tasks to each downstream \peis using set of virtual workers. 
%Moreover, \cg enables handling mutli-dimensional resource requirements, e.g., CPU and memory, in an efficient manner.
%\cg achieves the fair assignment of tasks by moving a virtual worker from the low capacity worker to the high capacity worker.
%Experiments show that \cg outperforms other approaches in terms of throughput and latency while providing significant improvement in terms of imbalance.
%The simple extension of the proposed approach for various use cases shows the flexibility and strength of the solution.

% !TEX root =  ../main.tex

%%% LITERATURE
\section{Background on Stream Partitioning}\label{sec:rel-work}

Load Balancing is one of the very well-studied problems in distributed systems.
Also, it is very extensively studied in theoretical computer science \cite{mitzenmacher2001potc-survey}.
%We refer to section \ref{sec:background} for  load balancing in stream processing systems and
%Next, we provide a brief overview of the load balancing problem for several large-scale distributed systems.
Next, we provide a discussion on various ways load balancing has been addressed in distributed systems, as well as state-of-art partitioning strategies to assign load to workers in such systems.

\subsection{Load Balancing in Distributed Systems}

%\spara{Graph processing systems.}
In graph processing systems, load balancing is often found along with balancing graph partitioning, where the goal often is to minimize edge-cut between different partitions \cite{gonzalez2012powergraph, malewicz2010pregel}.
Further, several systems have been proposed specifically to solve the load balancing problem, e.g., Mizan~\cite{khayyat2013mizan}, GPS~\cite{salihoglu2013gps}, and others.
Most of these systems perform dynamic load rebalancing at runtime via vertex migration \cite{yan2015effective}. % proposed a mirroring of high-degree vertices the to achieve better load balancing in pregel-like systems.

%\spara{Map-Reduce like systems.}
Load balancing and scheduling often appears in a similar context in map-reduce like systems, where the goal is to schedule the jobs to set of machines in order to maximize the resource utilization \cite{hindman2011mesos, vavilapalli2013apache}.
Sparrow~\cite{ousterhout2013sparrow} is a stateless distributed job scheduler that exploits a variant of the power of two choices~\citep{park2011multiplechoices}. 
\citet{ahmad2012tarazu} improves the load balance for map-reduce in heterogenous environment by monitoring and scheduling the jobs based on communication patterns.
%SkewTune~\cite{kwon2012skewtune} solves the problem of load balancing in MapReduce-like systems by identifying and redistributing the unprocessed data from the stragglers to other workers.

%\spara{Other distributed systems.}
Dynamic Load balancing in database systems is often implemented using rebalancing, similar to all the other systems  \cite{rahm1995dynamic}. 
Also, online load migration is effective for elasticity in the database systems \cite{suresh2015c3,2014marcoestore}. 
%\citet{serafini2016clay} propose an online partitioning approach that relies on identification of hot tuple from the loaded partition for migration along with it's co-accessed tuples.
Lastly, dynamic load balancing is considered in the context of web servers \cite{cardellini1999dynamic}, GPU \cite{chen2010dynamic}, and many others.

% !TEX root =  ../main.tex

%\subsection{Background}
\label{sec:background}
%In this section, we provide the brief summary of the state-of-the-art streaming solutions and discuss other possible solutions for our problem.

\subsection{Existing Stream Partitioning Functions}
\label{sec:existing-partitioning}
%% stream grouping
Messages are sent between \peis by exchanging messages over the network.
Several primitives are offered by \dspes for sources to partition the stream, i.e., to route messages to different workers.
%There are three main primitives of interest: \emph{key grouping}, \emph{partial key grouping} and \emph{shuffle grouping}.
%While these primitives receive different names in different platforms (e.g., key grouping is called ``fields grouping'' in Storm), they provide the same functionality.

%% key grouping
\spara{Key Grouping (\kg).}
This partitioning ensures the messages with the same key are handled by the same \pei (analogous to MapReduce).
It is usually implemented through hashing. %, i.e., hash(\key) $\mod \mid\workers\mid$.
\kg is the perfect choice for \emph{stateful} operators.
It allows each source \pei to route each message solely via its key,  without the need to keep any state or to coordinate among \peis.
However, \kg does not take into account the underlying skewness in the input distribution, which causes a few \peis to sustain a significantly higher load than others.
This suboptimal load balancing leads to poor resource utilization and inefficiency.

\spara{Partial Key Grouping (\pkgs).}
\pkgs \cite{nasir2015power, nasir2016two, nasir2015partial} adapts to the traditional power of two choices for load balancing in map-reduce like streaming operators.
\pkgs guarantees nearly perfect load balance in the presence of bounded skew using two novel schemes: key splitting and local load estimation.
The local load estimation enables each source to predict the load of workers leveraging the past history. 
However, similar to \kg, \pkgs assumes that each worker has the same resources and the service time for the messages follows a uniform distribution, which is a strong assumption of many real-world use cases. %, as discussed previously.

%% shuffle grouping
\spara{Shuffle Grouping (\sg).}
This partitioning forwards messages typically in a round-robin fashion.
It provides excellent load balance by assigning an almost equal number of messages to each \pei.
However, no guarantee is made on the partitioning of the key space, as each occurrence of a key can be assigned to any \pei.
It is the perfect choice for \emph{stateless} operators.
However, with \emph{stateful} operators one has to handle, store and aggregate multiple partial results for the same key, thus incurring additional memory and communication costs. 

\subsection{Consistent Hashing}

Consistent Hashing is a special form of a hash function that requires minimal changes as the range of the function changes \cite{karger1997consistent}.
%It was initially designed for caching, and was later adapted in variety of distributed systems like, Chord \cite{stoica2003chord}, Dynamo \cite{decandia2007dynamo}, Cassandra \cite{lakshman2010cassandra}, Memcache 
%\cite{nishtala2013scaling} and many others.
This strategy solves the assignment problem by systematically producing a random allocation.
It relies on a standard hash function that maps both messages and workers into unit-size circular ID space, i.e., $[0,1) \subseteq \real $.
Further, each task is assigned to the first worker that is encountered by moving in the clockwise direction on the unit circle.
%For implementation, the hash value for all the workers are stored in a binary search tree, and the clockwise successor can be found via single search.
Consistent Hashing provides load balancing guarantees across the set of workers.
Assuming $n$ are the number of available workers, and given that the load on a node is proportional to the size of the interval it owns, no worker owns more than $\BigO{\frac{\log n}{n}}$ of the interval (to which each task is mapped), with high probability \cite{karger1997consistent}.

One common solution to improve the load balance is to introduce \emph{virtual workers}, which are copies of workers, corresponding to points in the circle. 
Whenever, a new worker is added, a fixed number of virtual workers is also created in the circle.
As each worker is responsible for an interval on the unit circle, creating virtual workers spreads the workload for each worker across the unit circle.

Similar to other stream partitioning functions, consistent hashing does not take into account neither the heterogeneity in the cluster or the skewness in the input stream, which restricts its immediate applicability in the streaming context. 
A way to deal with both heterogeneity and skewness is to employ hash space adjustment for consistent hashing \cite{hwang2013adaptive}.
Such schemes require global knowledge of the tasks assigned to each worker to adjust the hash space for the workers, i.e., movement of tasks from the overloaded worker to the least loaded worker. 
Even though such schemes provide efficient results in terms of load balance, their applicability in stream context incurs additional overhead due to many-to-many communication across workers.
On the other hand, if implemented without global information, these schemes may produce unpredictable imbalance due to random task movement across workers.
%\notes{why random? please explain or fix the argument}

\spara{Consistent Hashing with Bounded Load (\ch).}
Independent from our work, \citet{mirrokni2016consistent} proposed a novel version of consistent hashing scheme that provides a constant bound on the load of the maximum loaded worker. 
The basic idea behind their scheme is to introduce the notion of capacity for each worker. 
In particular, set the capacity of each bin to the average load times ($1+\epsilon$), for some parameter $\epsilon$.
Further, the tasks are assigned to workers in the clockwise direction with spare capacity.
%\ch guarantees that the load of the maximum loaded bin is at most (1+$\epsilon$) factor of the average load. 
%In this paper, we randomly select the value of epsilon equals to \num{0.3}.
%\ch with bounded load can be adapted for our problem in a way that each message is assigned to the virtual bin in the clockwise direction. 

\subsection{Other Approaches}

\spara{Power of Two Choices (\potc).}
\potc achieves near perfect load balance by first selecting two bins uniformly at random and later assigning the message to the least loaded of the two bins. 
For \potc, the load of each bin is solely based on the number of messages. % as the information related to the service time is not available at upstream operators. 
Using \potc, each key might be assigned to any of the workers.
Therefore, the memory requirement in worst case is proportional to the number of workers, i.e., every key appearing on all the workers.

%\spara{Greedy Scheduling.}
%Sparrow~\cite{ousterhout2013sparrow} is a stateless distributed job scheduler that exploits a variant of the power of two choices~\citep{park2011multiplechoices}. 
%It employs batch probing, along with late binding, to assign $m$ messages of a job to the least loaded of $d \times m$ randomly selected workers ($d \geq 1$). 
%The applicability of such schemes in the context of streaming is not clear as both probing and late binding can significantly affect the latency per message.

\spara{Rebalancing.}
Another way to achieve fair assignment is to leverage \emph{rebalancing}~\citep{shah2003flux,balkesen2013adaptive,castro2013integrating, suresh2015c3, das2014adaptive}.
Once load imbalance is detected, the system activates a rebalancing routine that moves some of the messages and the state associated with them, away from an overloaded worker.
While this solution is easy to understand, its applicability in the streaming context requires answering challenging questions: How to identify the imbalance and how to plan the migration. % (how to keep ownership of the work), and (3) How to find the state of the messages after migration.
The answers to these questions are often application-specific, as they involve a trade-off between imbalance and rebalancing cost that depends on the size of the state to migrate. 
%Moreover, a major problem with moving messages explicitly is that the \dspes must maintain several routing tables: one for each worker so that clients can find out where to find the assigned messages.
%Each routing table has one entry for each task.
%Keeping these tables is impractical because of both search time and memory requirements.
%In a typical web mining application, each routing table can easily have billions of keys.
%Also, modifications to the routing table must be consistent across the workers, which creates further overhead.
For these reasons, rebalancing creates a difficult engineering challenge, which we address in our paper.

% !TEX root =  ../main.tex
%% PRELIMINARIES
\section{Preliminaries \& Problem Definition}
\label{sec:preliminaries}

This section introduces the preliminaries that are used in the rest of the paper.
%Also, we provide the brief summary of the state-of-the-art streaming solutions and discuss other possible solutions for our problem.

We consider a \dspe running on a cluster of machines that communicate by exchanging messages following the flow of a \dagr.
For scalability, streams are partitioned into sub-streams and processed in parallel on a replica of the \pe called \emph{processing element instance} (\pei).
Load balancing across the whole \dagr is achieved by balancing along each edge independently.
Each edge represents a single stream of data, along with its partitioning scheme.
Given a stream under consideration, let \sources be the set of sources, \workers be the set of workers, and their sizes be $|\sources| = \numsources$ and $|\workers| = \numworkers$.

Each \pei $\worker \in \workers$ is deployed on a machine  with a limited capacity $\capacity{\worker} \in \calC$.
For simplicity, we assume that there is a single important resource on which machines are constrained, such as storage and processing. 
Moreover, each \pei ($\worker \in \workers$) has an unbounded input queue (\workerqueue{w}).
%Capacities are normalized so that the average capacity is $\frac{1}{\numworkers}$; that is $\sum_{\worker \in \workers} \capacity{\worker}=1$.
%We assume them ordered by decreasing capacities, i.e., $c_1 \ge c_2 \ge c_3 \ldots \ge c_{\numworkers}$. 

The input to the \dagr is a sequence of messages $z = \tuple{\messageidentifier,\key,v, t_{\messageidentifier}}$ where $\messageidentifier$ is the identifier, $\key \in \keyspace$ is the message key, $v$ is the value, and $t_{\messageidentifier}$ is the timestamp at which the message is received.
The messages are presented to the engine in ascending order by timestamp.
Upon receiving a message with key $\key\in \keyspace$, we need to decide its placement  among the workers. 
We assume one message arrives per unit of time.

We employ queuing theory as the cost model to define the delay and the overhead at each worker.
In the model, a sequence of messages arrives at a worker $\worker \in \workers$. 
If the worker is occupied, the new message remains in the queue until it can be served.
After the message is processed, it leaves the system.
We represent the finish time for a message $\messageidentifier$ using \finishtime{\messageidentifier}.
The difference between the arrival time and the \finishtime{\messageidentifier} represents the latency of executing the message.

We define a {\em partitioning} function $\partitioner: \mathcal{\keyspace} \rightarrow \mathbb{\workers}$, which maps each message into one of the \peis.
This function identifies the \pei responsible for processing the message.
Each \pei is associated with one or more keys.
The goal of the partitioning function is to generate an assignment of messages to the set of workers in a way that average waiting time is minimized.

\begin{figure}[t]
\begin{center}
\includegraphics[width=0.8\columnwidth]{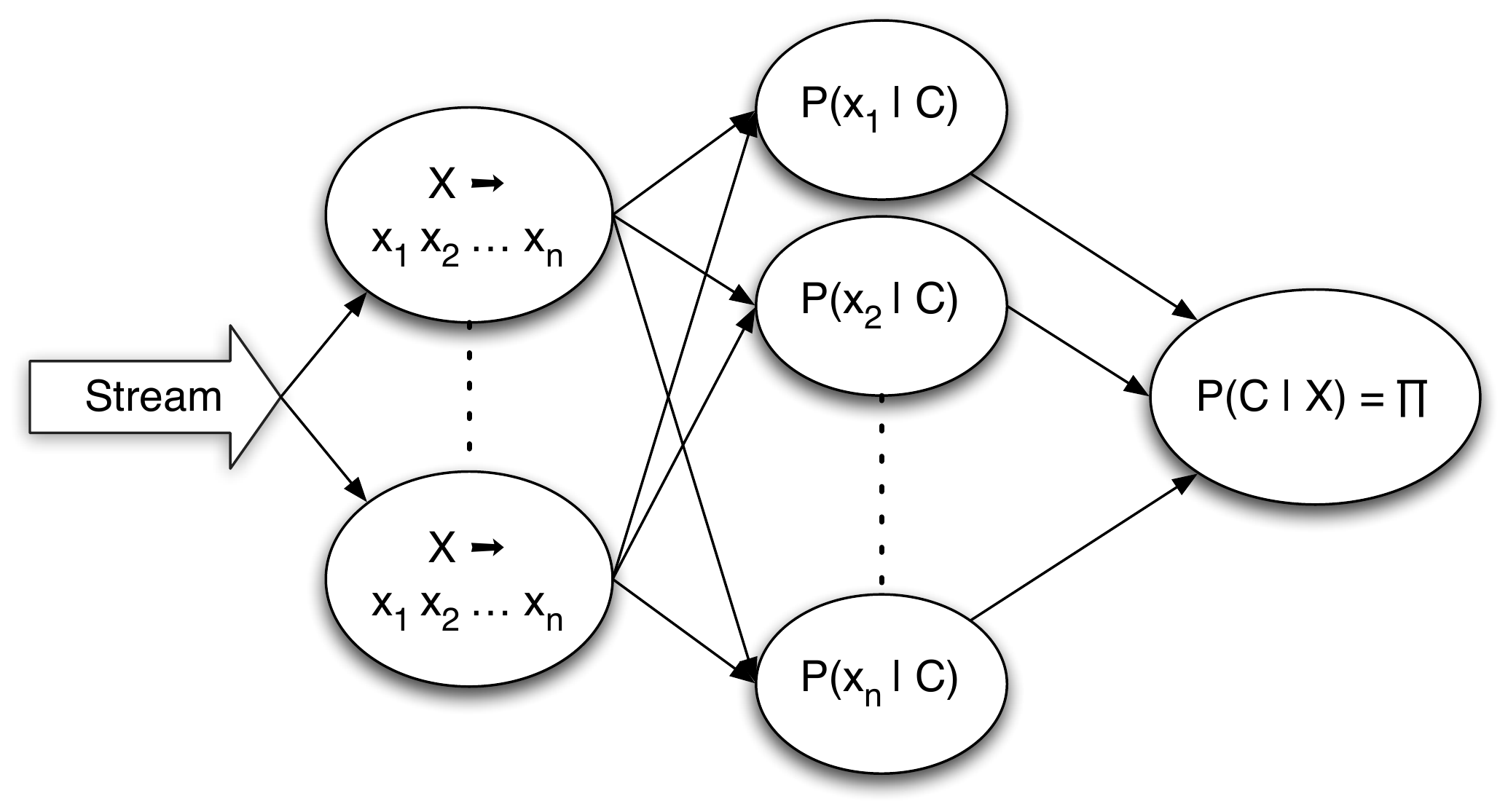}
\caption{Na\"{i}ve Bayes implemented via key grouping (\kg).}
\label{fig:vnb}
\end{center}
\vspace{-\baselineskip}
\end{figure}

%In this paper, we focus on providing a generalized framework for load balancing that is capable for adapting to any definition of load imbalance. 
%We define the \emph{queue length} of a worker using the number of messages that are pending in the queue.
%At time $t$, the \emph{queue length} of a worker \worker is defined by:
%$$ \workerqueue{w}({t}) = | { \{ \messageidentifier : (\partitioner(\messageidentifier,j)= \worker) \wedge (\finishtime{\messageidentifier} > t) \} }|, \text{ for } \worker \in \workers$$

We define the \emph{load} of a worker using the number of messages that are assigned to the worker at time $t$:
$$\load{\worker}{t}  = | { \{ \messageidentifier : (\partitioner(\messageidentifier,j)= \worker) \wedge (t_{\messageidentifier} < t) \} }|, \text{ for } \worker \in \workers$$

Also, we define {\em normalized load} at time $t$ as the ratio between the \emph{load} and the capacity of the worker.
$$ \utilization{\worker}{t} = \dfrac{\load{\worker}{t}}{\capacity{w}}$$

We use a definition of \emph{imbalance} similar to others in the literature (e.g., Flux~\citep{shah2003flux} and~\pkgs~\cite{nasir2015power}).
We define {\em imbalance} at time $t$ as the difference between the maximum and the average normalized load:
$$ I(t) = \max_{w}\{\utilization{\worker}{t}\} - \avg_{w}\{\utilization{\worker}{t} \}, \quad w \in \workers .$$

Further, the memory footprint of a worker $\worker$ is the number of unique keys assigned to the worker:
$$ M_{\worker}(t) = |\{ k_i : (\partitioner(\messageidentifier,j) = \worker) \wedge (\finishtime{\messageidentifier} < t)\}|, \text{ for } \worker \in \workers$$

\spara{Problem.}
Given the definition of imbalance, we consider the following problem in this paper.
\begin{problem}
 Given a stream of messages drawn from a heavy-tailed distribution $\keyspace$ and a set of workers $\worker \in \workers$ with capacities $\capacity{\worker} \in \calC$, find a partitioning function $\partitioner$ that minimizes memory footprint while keeping the imbalance ($ I(t)$) bounded by a constant factor at any time instance $t$.
\label{prob: 1}
\end{problem}

\spara{Memory Cost.} One simple solution to address problem \ref{prob: 1} is to employ round robin assignment as in \sg, which provides an imbalance of at most one in case of a homogenous cluster.
%Round robin achieves perfect load balance by assigning messages in a round robin fashion.
This load balance comes at the cost of memory, as messages with the same key might end up on all the workers. 
Also, the round robin assignment produces a higher aggregation cost \cite{nasir2015power, nasir2016two,katsipoulakis2017holistic}, which represents the communication cost for accumulating the partial results from the set of workers.
%These issues have been addressed in \cite{nasir2015power, nasir2016two, katsipoulakis2017holistic} and have been well received by both research and open source community.
%\notes{Have we talked about aggregation costs in those references? I don't know if I would say this sentence}

\spara{Example.}
\label{ex: naive}
To make the discussion more concrete, we introduce a simple application that will be our running example: the \emph{na\"{i}ve Bayes classifier}.
A na\"{i}ve Bayes classifier is a probabilistic model that assumes independence of features in the data (hence the na\"{i}ve).
It estimates the probability of a class $C$ given a feature vector $X$ by using Bayes' theorem:
$$
P(C | X) = \frac{ P(X | C) P(C) }{ P(X) }.
$$
The answer given by the classifier is then the class with maximum likelihood
$$
C^* = \argmax_C P(C | X).
$$
Given that features are assumed independent, the joint probability of the features is the product of the probability of each feature. Also, we are only interested in the class that maximizes the likelihood, so we can omit $P(X)$ from the maximization as it is constant.
The class probability is proportional to the product
$$
P(C | X) \propto \prod_{x_i \in X} P(x_i | C) P(C),
$$
which reduces the problem to estimating the probability of each feature value $x_i$ given a class $C$, and a prior for each class $C$.
In practice, the classifier estimates the probabilities by counting the frequency of co-occurrence of each feature and class value.
Therefore, it can be implemented by a set of counters, one for each pair of feature value and class value.
%A MapReduce implementation is straightforward, and available in Apache Mahout.\footnote{\url{https://mahout.apache.org/users/classification/bayesian.html}}
%This application is an adaptation of the classical MapReduce word count to the streaming paradigm where we want to generate a list of top-k words by frequency at periodic intervals (e.g., each $T$ seconds).
%It is also a common application in many domains, for example to identify trending topics in a stream of tweets.

%\spara{Design Goals.}
%%\dspe allows deployment of streaming applications that are represented using a \dagr.
%%The input in the system arrives in the form of an unbounded stream and various transformations are applied on the fly.
%%For scalability, streams are partitioned into sub-streams and processed in parallel on a replica of the \pe called \emph{processing element instance} (\pei).
%%Akin to Map-Reduce, grouping strategies are employed for assignment of messages to each \pei.
%\dspe requires supporting streaming applications with execution latencies as low as microseconds and throughput upto millions of messages per second. 
%Efficient resource utilization and fair scheduling are very crucial for achieving high throughput and low latencies.
%Due to such strict requirements on processing time, many solutions from other domains, like probing and solving bin packing problem cannot be directly adapted for streaming settings.

% !TEX root =  ../main.tex

%% PITCH
\section{Solution Primitives}
\label{sec:sol}
In this section, we discuss our solution and its various components.
Given a set of sources and a set of workers, the goal is to design a grouping strategy that is capable of assigning the messages to the workers proportionally to their capacity, while dealing with the messages' embedded skew.

\spara{Overview.}
In our work, we propose a novel grouping scheme called consistent grouping (\cg).
Our scheme borrows the concept of virtual workers from the traditional consistent hashing \cite{godfrey2004load,godfrey2005heterogeneity} and employs rebalancing to achieve fair assignment, similar to \citep{shah2003flux,gedik2014partitioning,balkesen2013adaptive,castro2013integrating, suresh2015c3}.
\cg allows variable number of  \emph{virtual workers} for each \pei.
The main idea behind \cg is to assign the input stream to the virtual workers in a way that each virtual worker has approximately equal number of messages.
Later, these virtual workers are assigned to the actual workers based on their capacity.
%The main idea of our algorithm is to divide the workload into small and equal-sized bins.
%Later, assign the number of bins to the workers based on their capacity.
%We employ traditional consistent hashing (\ch) along with virtual workers to achieve this goal, i.e., each virtual worker corresponds to a small bin.
One of the challenges is to bound the load of each virtual worker, as it implies that moving a virtual worker from one worker to another actually increases the receiving worker's load. %  increasing and decreasing the workload. % we move a bin from one worker to another, it
For this, we propose a novel grouping strategy called \emph{power of random choices} (\porc) that is capable of providing bounded imbalance while keeping the memory cost low. % balanced allocation of messages to the bins.
Further, we propose three efficient schemes within \cg: \emph{worker delegation}, \emph{piggybacking} and \emph{periodic aggregation}, which enable efficient integration of our proposed scheme into standard \dspes.
%Results show bla bla bla. % using traditional power of two choices \cite{azar1999balanced-allocations}, which ensures that the load of each bin is within $\BigO{\log{n}}$ factor from other bins.
Consistent grouping follows the same programming primitive as \pkgs for stream partitioning. % and not every algorithm can be expressed with it.
%In general, all algorithms that use shuffle grouping can use \cg to reduce their memory footprint.
%In addition, many algorithms expressed via key grouping can be rewritten to use \cg in order to get better load balancing.
We refer to \cite{nasir2015power} for the examples of common data mining algorithms that benefit from \cg. % , and show the advantages of \cg. % in terms of load 

\subsection{Power of Random Choices}
\label{sec:bal-all}
%In this section, we explain \porc. %  and briefly discuss several other grouping schemes for comparison. %Next, we propose \emph{power of a random choice}, which 
%\porc is a hybrid scheme between \potc and \pkgs.
\porc assigns the incoming messages to the set of virtual workers in a way that the imbalance is bounded and the overall memory footprint of the keys on the virtual workers is low.
The basic idea behind \porc is to introduce the notion of continuous capacity, which is a function of average load. 
In particular, we set the capacity of each virtual worker to the average load times ($1+\epsilon$), for some parameter $\epsilon$.
Note that the definition of capacity is based on the average load, rather than a hard constraint. 
Given a sequence of virtual workers for a key, \porc maps the key to the first virtual worker with the spare capacity.
The sequence of virtual workers for a key are produced by using a single hash function and concatenating the \textit{salt} with the key to produce a new assignment\footnote{https://datarus.wordpress.com/2015/05/04/fighting-the-skew-in-spark/}.
We refer to the first virtual worker in the sequence as the \emph{principal virtual worker}.
The rational behind this approach is that the heavy keys in the skewed input distribution overload their principal worker.
Therefore, we allow the heavy keys to spread across the other virtual workers, which reduces the memory footprint compared to other schemes, e.g., round robin.
The $\epsilon$ parameter in \porc provides the trade off between the imbalance and memory footprint.
Algorithm \ref{alg: porc} provides the pseudocode.
\porc provides an efficient and generalized solution for the fundamental problem of load balancing for the skewed stream in streaming settings, while minimizing the memory footprint. % \cite{nasir2015power,nasir2016two}.
%\notes{Why is that? We never talked about PoRC and its memory footprint in those references. please fix}
In our work, we adapt \porc for fair load balancing for streaming applications, which shows its effectiveness and applicability.
%Independent from our work, \citet{mirrokni2016consistent} proposed a similar strategy to bound the maximum load among the set of workers using consistent hashing. 
%We refer to their scheme as \ch throughout our paper.

\begin{algorithm}
\footnotesize
  \caption{Pseudocode for Power of Random Choices.}
\label{alg: porc}
  \begin{algorithmic}[1]
  \Require key, hash-function, $\#$messages, $\#$workers, load-vector, imbalance-factor
  \Ensure $S^*\in\{1\ldots n\}$
 \Procedure{getWorker}{$\key$, $\hashk$ , $m_t$ , $\numworkers$, $load$, $\epsilon$}   
	\State $salt \gets 1$
	\State $S^* \gets \hashf{\key+salt}$
 	\While{($load[S^*] \ge (1+\epsilon) \, \frac{m_t}{\numworkers}$)} 
 		\State $salt \gets salt+1$
 		\State $S^* \gets \hashf{\key+salt}$
	\EndWhile
	\State $load[S^*]\gets load[S^*]+1$
	\State \Return $S^*$
\EndProcedure

%\notes{What is ti doing here? Please check}
 \end{algorithmic}
\end{algorithm}

\begin{figure}[t]
\begin{center}
\includegraphics[width=\columnwidth]{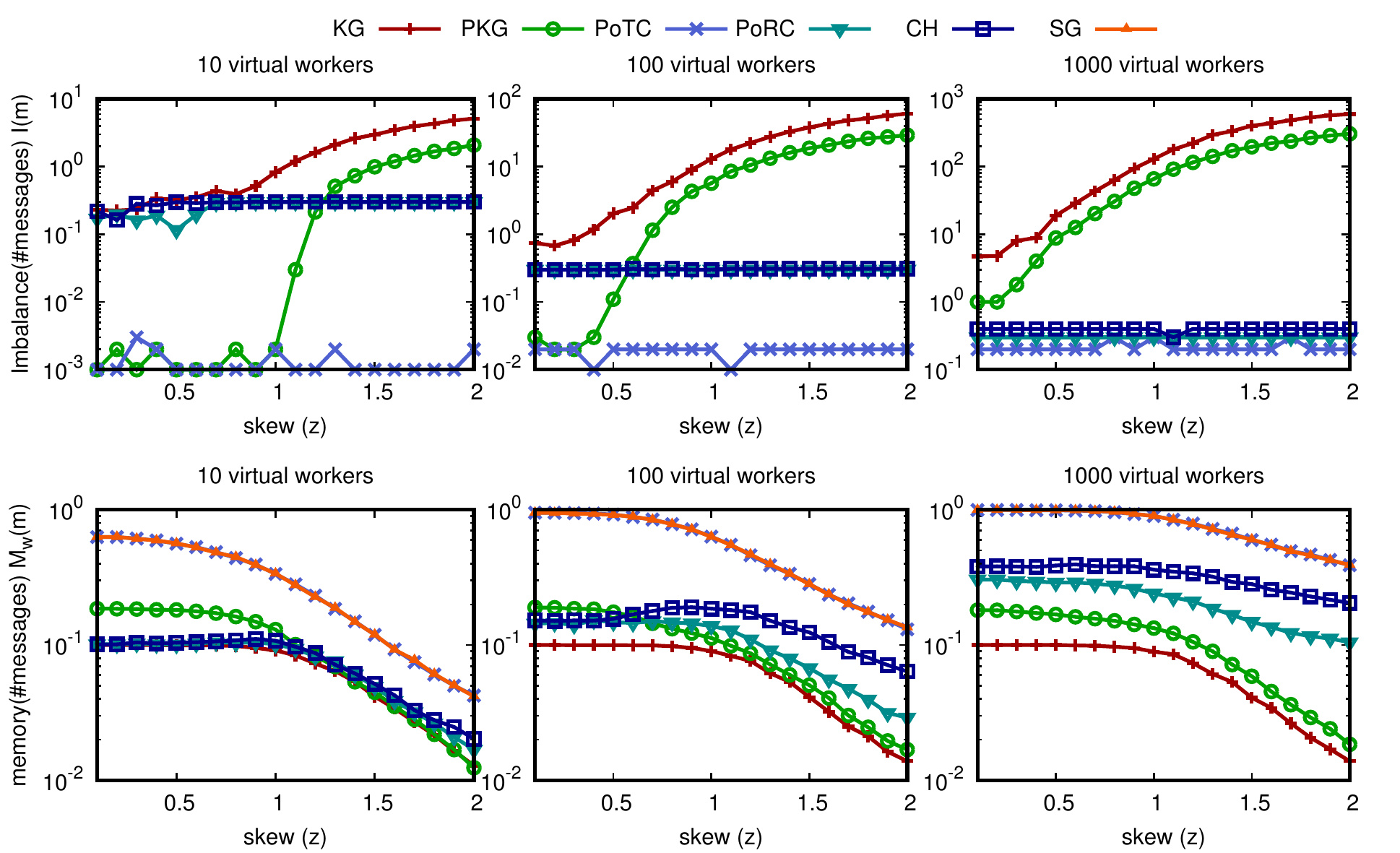}
\caption{Normalized imbalance and memory overhead for different schemes using zipf distribution with different skew and number of virtual workers.
\label{fig:imbalance-balls-and-bins}
%\notes{The memory in y-axis is not I(t) right? is something else? Or are you referring to imbalance? Also please find a way to show \sg. maybe by reordering how you plot them it will show?}
}
\end{center}
\vspace{-\baselineskip}
\end{figure}

%\begin{table}[t]
%\caption{Normalized imbalance when varying the number of virtual workers for the Wikipedia (WP) and Twitter (TW) datasets.
%We set the value of $\epsilon = 0.3$ for \potc and \ch.}
%%\vspace{-2mm}
%\tabcolsep=0.1cm
%\centering
%\small
%\begin{tabular}{l c c c c c c c c}
%\toprule
%Dataset		&	\multicolumn{4}{c}{WP} 					&	\multicolumn{4}{c}{TW}					\\
%\cmidrule(lr){2-5} \cmidrule(lr){6-9}
%$W$			&	$10^1$		&	$10^2$		&	$10^3$		&	$10^4$		&	$10^1$		&	$10^2$		&	$10^3$		&	$10^4$		\\
%\midrule
%\kg		&		\num{0.8}	&	\num{9.15}		&	\num{93}	&	 \num{931}	&		\num{2.2}	&	\num{25}		&	\num{246}	&	\num{2469}	\\
%\pkgs		&		8e-7	&		3.67	&	45.59	&	464	&	\num{1.52}		&	\num{11.34}		&	\num{22}	&	\num{1233}	\\
%\potc		&		8e-7	&		8e-6	&	9e-5	&	1e-3	&	9.5e-8		& 2e-6		&	2e-5	&	3e-4	\\
%\sg		&		4e-7	&	3e-6		&	4e-6	&	2.8e-4	&	\num{0}		&	2e-7		&	6e-6	&	4e-5	\\
%\porc		&	\num{0.3}	&	\num{0.3}	&	\num{0.3}	&	\num{0.3}	&	\num{0.3}	&	\num{0.3}	&	\num{0.3}	&	\num{0.3}	\\
%\ch		&		\num{0.3}	&	\num{0.3}		&	\num{0.3}&	\num{0.3}	&	\num{0.3}		&	\num{0.3}			&	\num{0.3}	&	\num{0.3}		\\
%\bottomrule
%\end{tabular}
%\label{tab:bmf}
%%\vspace{-3mm}
%\end{table}

\spara{Discussion.}
%Note that all the aforementioned schemes are lightweight and only require constant $\BigO{1}$ time to produce assignments.
To show the effectiveness of \porc, we compare its performance with \kg, \pkgs, \potc, \sg, and \ch \cite{mirrokni2016consistent} in terms of imbalance and memory footprint (see section \ref{sec:background} for the description of these schemes).
We leverage a zipf-based dataset with different skews for this experiment (see Section~\ref{sec:eval} for the description of the dataset).
The top row of Figure \ref{fig:imbalance-balls-and-bins} reports the imbalance for different schemes for different number of virtual workers, i.e., $10$, $100$, and $1000$. %comparison of these schemes in terms of imbalance.
Results show that both key grouping and partial key grouping generate high imbalance as the skew and the number of virtual workers increase. 
However, the other schemes perform fairly well in terms of imbalance. % generate low imbalance. 
%Similarly, Table \ref{tab:bmf} shows the imbalance for the Wikipedia and Twitter dataset when varying the number of virtual workers.
%The load balance is achieved at the cost of memory \cite{nasir2016two} and any sophisticated assignment scheme  \cite{nasir2016two} can replace these strategies to reduce the memory footprint.
Additionally, we report the memory overhead for all the schemes in Figure \ref{fig:imbalance-balls-and-bins}.
The memory cost is calculated using the total number of unique keys that appear at each virtual worker.
Results verify our claim that load balance is achieved at the cost of memory.
%Based on aforementioned discussion, we rely on power of two choices henceforth.
%Further, we define consistent grouping, which relies on \porc to solve the problem \ref{prob: 1}.
%\notes{I don't get this line}

\subsection{Consistent Grouping}
We propose a novel grouping strategy called \emph{Consistent Grouping} (\cg), inspired by consistent hashing.
\cg borrows the concept of virtual workers from traditional consistent hashing and allows variable number of virtual workers for each \pei \cite{godfrey2004load, godfrey2005heterogeneity}. 
It is a dynamic grouping strategy that is capable of handling both the heterogeneity in the resources and the variability in the input stream at runtime.
\cg achieves its goal by allowing the powerful workers to acquire additional virtual workers, which leads to `stealing' work from the other workers.
Moreover, it allows overloaded workers to gracefully revoke some of their existing virtual workers, which is equivalent to giving up on some of the allocated work.

\cg is a lightweight and distributed scheme that allows assignment of messages to the workers in a streaming fashion.
Moreover, it leverages \porc for assignment of keys to each virtual worker in a balanced manner, which allows it to bound the load of each virtual worker.
%As \cg adapts to consistent hashing, it inherits the basic benefits of \ch, i.e., supporting search operation on $n$ workers in $\mathcal{O}(\log n)$ time and efficiently dealing with systems where the set of workers available varies over time.
\cg is able to balance the load across workers based on their capacities, which allows the \dspe to operate under realistic scenarios like heterogeneous clusters and variable workloads.

\spara{Time Slot.}
We introduce the notion of \emph{time slot} ($t_0$), which represents the minimum monitoring time period for a \pei. % waits after sending a signal to the upstream \peis. 
$t_o$ is an administrative preference that can be determined based on workload traffic patterns.
%Typically, only a relatively small portion of the hash space controllable by a second system parameter is rearranged during each scheduling event. 
If workloads are expected to change on an hourly basis, setting $t_0$ on the order of minutes will typically suffice.
For slower changing workloads $t_0$ can be set to an hour.
Time slot guarantees that workers have enough sample of the input stream to predict their workload. % continuously from the same downstream \pei.

%\subsubsection{Load Reduction.}
Similar to consistent hashing, \cg initializes with the same number of virtual workers for each worker, i.e., $\BigO{\log {n}}$.
\cg manages a unit-size circular ID space, i.e., $[0,1) \subseteq \real $, and maps the virtual workers and keys on the unit-size ID space. 
%The main idea is to assign lower number of virtual workers for low capacity workers and higher number of virtual workers for high capacity workers.
%However, it is not clear how \cg can reassign the keys or move the virtual workers ensuring that the load of each worker is proportional to its capacity.
We would like a scheme that is capable of monitoring the load at each worker throughout the lifetime of a streaming application and adjust the load according to the available capacity of the workers.
%For simplicity, assume that upstream operators know that one of the worker is overloaded (we will remove this assumption in the later sections).
%\cg requires reducing the load of the overloaded worker.
In doing so, we introduce a novel scheme called \emph{pairing virtual worker.} %to reduce the load of a worker:
%\notes{Not sure what these two paragraphs contributed. Please make more explicit later why we need this time slot. Also, if this is actually a time period and not a time instance, then define it as dt or $\Delta$t and consistently from here on}

\spara{Pairing virtual workers.}
The load of a worker equals to the sum of loads of the assigned virtual workers.
Further, the load of each virtual worker equals to the load that is induced by the mapped messages. 
%Now a simple solution would be to randomly remove a virtual worker from one overloaded worker.
%However, this might produce unpredictable imbalance across the workers. %overload the other workers. %, as the neighboring worker will become responsible for the keys of the removed virtual worker.
Ideally, we would like to assign one of the virtual workers from the overloaded worker to one of idle workers.
However, it is not trivial  until this point on how one can achieve such an assignment. 
To enable such an assignment, we propose to maintain two first-come-first-serve (FCFS) queues: \emph{idle} and \emph{busy}.
These queues maintain the list of idle and busy workers in the \dspe and allow \cg to pair any removal and addition with the opposite to balance the number of virtual workers throughout the execution.
For instance, when a worker is overloaded, it sends a message to the sources. 
Further, the source only removes the virtual workers of the corresponding worker if it is able to pair it with an addition on an idle worker. %uniformly at random and spawn a new virtual worker for a worker in the FCFS queue.
This simple scheme ensures that the number of virtual workers in the system are balanced throughout the execution and the load of each virtual worker is bounded, which enables \cg to perform fair assignment.
Note that mapping the virtual workers with similar keys to the same worker might reduce the memory footprint. 
However, this requires maintaining all the unique keys in each virtual worker and each worker.
Therefore, we opt for FCFS mapping of virtual workers to workers.

%\spara{Shrink a virtual worker.}
%\label{sec:shrink}
%\porc independently ensures that the equal number of messages are assigned to each virtual worker.
%However, balancing the number of messages across the set of virtual workers does not ensure that the load of each virtual worker is same.
%This behavior is due to the fact that the service time for each key is different.
%Therefore, we need a scheme that can squeeze a virtual worker to ensure that the variance of the load of each virtual worker is low.
%Based on this observation, we propose a scheme that reduces the load of a virtual worker by dividing it into half.
%This simple scheme allows to reduce the load of a overloaded worker even in the case when there are no idle workers, as discussed in previous section.
%This scheme spreads the load equally among all the other virtual workers, hence not overloading any single virtual worker.

\subsection{Integration in a \dspe}

While consistent grouping is easy to understand, its applicability to the case of real world stream processing engines is not trivial. 
%In particular, we need to answer two questions: 1) How to identify the imbalance (what are the metrics for imbalance) and 2) how to plan the migration (how to keep ownership of the work).
%To answer these questions, 
We package \cg with few efficient strategies that enable its applicability in a variety of \dspes.

\spara{Worker Delegation.}
First, we propose an efficient scheme called \emph{worker delegation}, which pushes the load balancing problem to the workers and allows them to decide their workload based on their capacity.
%This simple modification reduces the distributed load balancing problem to a local load monitoring problem.
Each worker requires monitoring its workload and needs to take the decision based on their current workload and the available capacity.
The decision can either be to increase the workload or to decrease the workload.
The intuition behind this approach is that it is often the case that the cluster consists of a large number of workers and collecting the statistics periodically from the workers creates an additional overhead for a streaming application. 

The worker delegation scheme allows the workers to interact with sources by sending binary signals: (1) increase workload and (2) decrease workload.
Each worker monitors its workload and tries to keep the workload between two thresholds, i.e., if the workload exceeds the upper threshold, the worker sends a decrease signal to the sources, and if the workload is below the lower threshold, the worker sends the increase signal to the sources.
This simple modification comes along with the benefit that it gives the flexibility to the workers to easily adapt to the complex application-specific requirements, i.e., processing, storage, service time and queue length.

 \spara{Piggybacking.}  %For scalability, streams are partitioned into sub-streams and processed in parallel by replicating both sources and workers. 
%Considering the \dagr in the Figure \ref{fig:vnb}, we can observe that the performance of \cg is dependent on the parallelism of the operator.
%For instance, if the number of sources (upstream operators) are greater than the number of workers (downstream operators), \cg might negatively impact the performance of a streaming application.
Each worker requires updating all the sources in case of experiencing undesirable (low or high) workload.
Note that it is required that sources receive the signal and operate in a consistent manner, performing the same routing of messages.
Such deployment might negatively impact the performance of a streaming application, as it requires one-to-many broadcast messages across the network.
To overcome this challenge, we propose to relax the consistency constraint in the \dagr and allow operators to be eventually consistent.  
We propose to \emph{encode} the binary signals from the workers along with the acknowledgment messages. 
During the execution, the sources only receive the signal from the worker as a response to its messages.
This means that the worker might continue receiving the messages with the same key even after triggering the decision. 
%However, in worst case the number of messages received by downstream \peis is upper bounded by the number of upstream operators, assuming no network delays. 

\spara{Periodic Aggregation.}
When the sources receive a request to increase the workload, they move one of the virtual worker from the overloaded worker to an idle worker. % a virtual worker from a overload worker and a increases the workload of a worker by spawning new virtual workers.
%Similarly, when the upstream \pei receives a request to decrease the workload, it reduces the workload of a worker by removing one of the existing virtual workers.
During the change of routing, we need to ensure that the messages that are pending in the queue of the workers must be processed in a consistent manner.

%A way to handle such changes is to allow the newly spawned worker to steal the set of messages from the worker that appears next to it in the clockwise direction on the unit circle (successor).
%Similarly, the messages that were previously assigned to the removed worker must be moved to the worker responsible for the adjacent interval (successor).
%These two assignment operations are exactly similar to \ch.
%However, message migration over the network presents a difficult engineering challenge, as it requires moving the in-flight messages while the new messages are entering the \dagr \cite{castro2013integrating}.

\cg ensures that each message is processed in a consistent manner by discarding the message migration phase. 
%When the upstream \pei receives a request to change (increase or decrease) the workload, it moves the , which only affects the future routing of the messages.
Concretely, each worker processes the messages that are assigned to it, and any change in the routing only affects the messages that arrive later.

As a message with the same key might be forwarded to different workers, \cg performs \emph{periodic aggregation} of partial results from the workers to ensure that the state per key is consistent.
Periodic aggregation leverages the same \dagr and imposes a very low overhead in the stream application.
Particularly, \cg follows the same programming primitive as \pkgs for stream partitioning, supporting both stateless and stateful map-reduce like applications.
\section{Analysis}
\label{sec:theory}

We proceed to analyze the conditions under which \cg achieves good load balance.
Recall from Section \ref{sec:preliminaries} that we have a set \workers of \numworkers workers at our disposal.
Each worker $\worker \in \workers$ has a limited capacity, which is represented by $\capacity{\worker}\in \calC$.
Capacities are normalized so that the average capacity is $\frac{1}{\numworkers}$, that is $\sum_{\worker \in \workers} \capacity{\worker}=1$.
We assume that they are ordered in decreasing order of capacity, i.e., $c_1 \ge c_2 \ge c_3 \ldots \ge c_{\numworkers}$.
%For simplicity, we assume that there is a single important resource on which workers are constrained, such as storage, and processing.
%Moreover, each worker ($\worker \in \workers$) has an unbounded input queue (\workerqueue{w}).

The input to the engine is a sequence of messages $z = \tuple{\messageidentifier,\key,v, t_{\messageidentifier}}$ where $\messageidentifier$ is the identifier, $\key \in \keyspace $ is the message key, $v$ is the value, and $t_{\messageidentifier}$ is the timestamp at which the message is received.
Upon receiving a message with key $\key\in \keyspace$, we need to decide its placement
among the workers.
We assume one message arrives per unit of time.
The messages arrive in ascending order by timestamp.

\spara{Key distribution.}
We assume the existence of an underlying discrete distribution $\calD$ supported on $\keyspace$ from which keys
are drawn, i.e., $k_1, \ldots, k_{\keysize}$ is a sequence of $m$  independent samples from $\calD$ ($\keysize \gg \numworkers$).
%Without loss of generality, we identify the set $\keyspace$ of keys with
%$\naturals^+$ or, if $\keyspace$ is finite of cardinality $\keysize = |\keyspace|$, with $[\keysize ] = \{1, \ldots, \keysize \}$.
We represent the average arrival rate of messages as $p_{\key}$ and the cardinality of set \keyspace as \colors, i.e., $\colors = \mid \keyspace \mid$.
We assume that they are ordered in decreasing order of average arrival rate, %if $p_{\key}$ is the probability of drawing key $\key$ from $\calD$, then
$p_{1} \ge p_{2} \ldots \ge p_{\colors}$, and $\sum_{\key \in \keyspace} p_{\key} = 1$.
%The distribution of the arrival process is often modeled as a Poisson process \cite{harchol2000task}.
We model the load distribution as a zipf distribution with values of exponent $z$ between $0$ and $2.0$.
The probability mass function of the zipf distribution with $z$ is
$$f(r,\colors,z) = \frac {1/r^z }{\sum_{x=1}^{\colors} (1/x^z)},$$
where \emph{r} is the rank of each key, and $\keysize$ is the total number of elements.

Our goal is to design an algorithm to solve Problem \ref{prob: 1}.
%The problem we address is similar to the online bin packing problem; however, differs in the case that the number of bins are fixed and the capacities of the bins are not known in prior.
%In case of uniform distribution of the worker capacities and the load per key, equal number of keys can be assigned to each worker using either hash-based (key grouping) or round robin (shuffle grouping) scheme.
In the analysis of \cg, we assume that $t_0$ represents the time slot which corresponds to the minimum time period that each worker waits after sending a signal to the workers.
Also, as we are not considering elasticity, we assume that the system is well provisioned, i.e., $\frac{\sum_{\key \in \keyspace}p_{\key}}{\sum_{\worker \in \workers}\capacity{\worker}} < 1$.

%\notes{check if to is to be changed}

\subsection{Imbalance with Consistent Grouping}
For simplification, we divide the analysis of \cg into two parts: dividing the workload into small equal-sized virtual workers and assigning the virtual workers to workers based on their capacities.
Assume that $\alpha >1 $ represents the number of virtual workers assigned to each worker at initial time.
Then, for $n$ heterogeneous workers, we have $\alpha \times n$ homogeneous virtual workers.
Each virtual worker has the same capacity (hence, homogeneous), and the capacity is guaranteed to be at most the capacity of the worker with the lowest capacity.
The sources do not know the capacity of each worker.
However, since all virtual workers are homogeneous, the sources can balance the load of each worker
by assigning equal number of messages to each virtual worker, and by keeping the number of virtual workers
assigned to each worker proportional to its capacity.

\subsubsection{Chromatic Balls and Bins}
We model the first problem using the framework of balls and bins processes, where keys correspond to colors, messages to colored balls, and virtual workers to bins.
Choose $d$ independent hash
functions $\hash{1}, \ldots, \hash{d}\colon \calK \to [\alpha n]$ uniformly at random. %\footnote{\scriptsize In theory this requires $\Omega(|\calK| \log n)$ bits of storage, but practical implementations often use a PRG seed or a fixed hash function with a chosen ``salt'' string.}
Define the \code{greedy-$d$} scheme as follows:
at time $t$, the $t$-th ball (whose color is $k_t$) is placed on the bin with minimum current load among
$\hash{1}(k_t), \ldots, \hash{d}(k_t)$, i.e., $P_t(k_t) = \operatorname{argmin}_{i \in \{\hash{1}(k_t),\ldots,\hash{d}(k_t)\}} L_i(t)$.
We define the \emph{imbalance} as the difference between the maximum and the average load across the bins, at time $t$.
%, where ties are broken arbitrarily.
%Recall that with key splitting there is no need to remember the choice for the next time a ball of the same color appears.
%there is no need to stick to the same choice, although the set of $d$ candidate bins will be the same.)

Observe that when $d = 1$, each ball color is assigned to a unique bin so no choice has to be made; this models hash-based key grouping.
At the other extreme, when $d \gg n \ln n$, all $n$  bins are valid choices, and we obtain shuffle
grouping.% (the reason $d = n$ does not quite suffice for this claim is that the $d$ independent choices are not guaranteed to be different).
%Recall that in the standard balls and bins problem,
%the maximum load can be shown to be $O(m / n + \log n / \log \log n $O(m / n + \log \log n / \log d + 1)$ with high probability.

%Note that in case of shuffle grouping, balls with same color might be assigned to different workers.
%Such an assignment requires extra memory to gather partial states and requires an additional aggregate phase in case of stateful operators, e.g., aggregate, max, min.
%We express this behavior in terms of memory and aggregation cost.
%Further, observe that in case of key grouping, each ball is assigned to a single worker.
%Therefore, key grouping does not require any additional memory and the aggregation phase.

\pkgs \cite{nasir2015power} considers the case of $d = 2$, which is same as having two hash functions $\hash{1}(\key)$ and $\hash{2}(\key)$.
The algorithm maps each key to the sub-stream assigned to the least loaded worker between the two possible choices, that is:
$P_{t}(\key)$~=~$\argmin_i(L_i(t) : \hash{1}(\key) = i \lor \hash{2}(\key) = i)$.
%This simple modification allows to achieve an upper bound on the imbalance of $O(\log(\log(n)))$ where $n$ is the number of messages.
%As shown in Section~\ref{sec:evaluation}, this scheme actually achieves almost perfect load balance in our experiments.
\begin{lemma}\label{thm:main}
Suppose we use $n$ bins and let $m \ge n^2$.
Assume a key distribution $\calD$ with maximum probability $p_1 \le \frac{1}{5 n}$.
Then, the imbalance after $m$ steps of the \code{Greedy-$d$} process is $\BigO{\frac{\ln \ln n}{\ln d}}$, with high probability \cite{nasir2015power}.
\end{lemma}

%\notes{not sure if this should be called fact, or lemma, or theorem. Check the cited work and use the same}

Observe that the imbalance in case of \pkgs is only guaranteed for the case when $p_1 \le \frac{1}{5 n}$.
However, in the case when $p_1 > \frac{1}{5 n}$, the imbalance grows proportional with the frequency of the most frequent key and number of workers.

Power of Two Choices (\potc) \cite{azar1999balanced-allocations} leverages two random numbers $\mathcal{R}{1}(m)$ and $\mathcal{R}{2}(m)$.
The algorithm maps each message $m$ to the sub-stream assigned to the least loaded worker between the two possible choices, that is:
$P_{t}(k)$~=~$\argmin_i(L_i(t) : \mathcal{R}{1}(m) = i \lor \mathcal{R}{2}(m) = i)$. The above random numbers can be generated by using hash functions with messages as arguments. In this case, note that the \potc is different from the \pkgs in the sense that two hashes are applied to the messages, rather than the keys.
The procedure is identical to the standard \code{greedy-$d$} process of \citet{azar1999balanced-allocations}, therefore the following bounds hold.
%This simple modification allows to achieve an upper bound on the imbalance of $O(\log(\log(n)))$ where $n$ is the number of messages.
%As shown in Section~\ref{sec:evaluation}, this scheme actually achieves almost perfect load balance in our experiments.
\begin{lemma}\label{lem:potc}
Suppose we use $n$ bins and let $m \ge n^2$.
Then, the imbalance after $m$ steps of the \code{Greedy-$d$} process is $\BigO{\frac{\ln \ln n}{\ln d}}$, with high probability \cite{azar1999balanced-allocations}.
\end{lemma}
%\notes{not sure if this should be called fact, or lemma, or theorem. Check the cited work and use the same}

Note that these bounds can be generalized to the infinite process in which $n$ balls leave the system in each time unit (one from each worker) and the number of balls entering the system is less than $n$.
In such cases, the relative load remains the same, therefore the bound holds.
%Before proceeding to the next part of the analysis, we recall the more generalized theorem from \cite{berenbrink2000balanced}.
%\begin{fact}
%\label{fact:greedy-1}
%Suppose we allocate $m$ balls to $n$ bins using \code{Greedy-$d$} process with $d\geq 2$. Then the number of bins with load at least $\frac{m}{n}+i+\gamma$ is at most $n\cdot \exp(-d^i)$, with high probability, where $\gamma$ denotes a suitable constant.
%\end{fact}
%We refer to \cite{mirrokni2016consistent} for the analysis related to the imbalance in case of consistent hashing with bounded load.
%Moreover, we leave the analysis for \porc for future work and just show empirical results in this paper.
\porc generate imbalance that is bounded by the factor $\epsilon$ , i.e., $I(m) \leq \epsilon \cdot (\frac{m}{n})$.

\subsubsection{Fair Bin Assignment}

%\note{Anis- This section is incomplete}
Given that $m$ messages are assigned to set of $n$ workers using \porc, our goal is to show that consistent grouping is able to perform fair assignment to messages to the workers over time. %, one of the assignment scheme discussed in section \ref{sec:bal-all}.
%\ch manages a unit-size circular ID space, i.e., $[0,1) \subseteq \real $ employing arithmetic modulo 1.
%\cg manages a unit-size circular ID space, i.e., $[0,1) \subseteq \real $ to which both keys and workers are mapped.
%Both the workers ($\worker \in \workers$) and the keys ($\key \in \keyspace$) are mapped to the ID space using the hash function \partitioner.
%Each message with a key is hashed and assigned to the worker in the clockwise direction.
%The load of each worker is equal to the sum of load of the keys that reside in its ID space.
%Each worker may pick multiple IDs (virtual workers) to balance the workload across the set of workers.
%Our goal is to show that consistent grouping is able to perform fair assignment to messages to the workers over time.
We achieve our goal by showing that consistent grouping reduces the imbalance $I(t)$ (if it exists) over time.
To make the discussion more concrete, we define the notion of busy worker using a threshold $\theta _b > 1$.
In particular, we say that a worker $w$ is busy if the load $\staticload{\worker} \ge  \theta _b \cdot \capacity{\worker}$.
Similarly, we define the notion of idle worker using the threshold $\theta _i < 1$.
We say that a worker $w$ is idle if its load $\staticload{\worker} \le  \theta _i \cdot \capacity{\worker}$.

%\spara{Heterogeneous Cluster.}
Assume that $\alpha$ represents the average number of virtual workers per worker, i.e., the total number of virtual workers equal $\alpha \times n$.
Also, assume that $\alpha_{\worker}^*$ represents the optimal number of virtual workers for ${\worker}$-th worker, namely, $\alpha_{\worker}^* =  \capacity{\worker}n\alpha$. Clearly, $\frac{\theta _i \cdot \capacity{\worker}}{\alpha_{\worker}^*} \leq \frac{1}{n.\alpha} \leq \frac{\theta _b \cdot \capacity{\worker}}{\alpha_{\worker}^*}$.
%The expected load per virtual worker is $\BigO{\frac{1}{n\alpha}+\gamma}$, where $\gamma = O(1)$ represents the imbalance of \porc. %due to the assignment scheme, e.g., $\BigO{\frac{\log{n}}{n}}$ in case of hashing, $\BigO{\frac{\log{ \log {n} }}{\log{2}}}$ in case of \potc and $\BigO{1}$ for \ch and \porc.
%Now assume that

%Now, consider \cg instance in which each worker is initialized with the exactly same number of virtual workers.
%Each worker may pick multiple IDs (virtual workers) to balance the workload across the set of workers.
Thanks to  the load balancing mechanisms such as \pkgs or \potc, each virtual bin is guaranteed to
have load at most $1/(\alpha n) + \gamma$ with high probability, where $\gamma$ denotes the imbalance factor
of the load balancing mechanism used. For \pkgs and \potc, the value of $\gamma$
is at most $\left(\ln\ln{\alpha n}/(m \ln{d})\right)$ as implied by Lemmas~\ref{thm:main} and~\ref{lem:potc}
(notice that the denominator $m$ is due to the normalization of the capacity in this paper).
Therefore, the expected load of a worker $\worker$ having
$\alpha_{\worker}$ virtual workers is bounded above by%(using Fact \ref{fact:greedy-1}):
$$\mathbf{E}[\staticload{\worker}] \le \alpha_{\worker} \cdot (\frac{1}{n\alpha} +\gamma)$$

Now, consider that the worker $\worker$ is overloaded, i.e., $\mathbf{E}[L_\worker]\ge\theta _b \cdot \capacity{\worker}$.
This implies:
~$$\alpha_{\worker} \cdot (\frac{1}{n\alpha} +\gamma) \ge \theta _b \cdot \capacity{\worker}$$

We can rearrange the above equation to have:
$$\gamma \ge \frac{\theta _b \cdot \capacity{\worker}}{\alpha_{\worker} }- \frac{1}{n\alpha},$$
which implies that when the worker is overloaded, it must have an imbalance that is lower bounded by the
above equation. However, such an imbalance is guaranteed to be small $\epsilon$ by the load balancing mechanism
used, i.e., $\epsilon \le  \left(\ln\ln{\alpha n}/(m \ln{d})\right) \ll 1/(\alpha n)$ for $\potc$ and $\pkgs$, when $m \ge n^2$.

%\notes{what do you mean "be small $\epsilon$ by ..." Please explain}

Therefore, we know that for an overloaded worker, it must hold that:
$$\epsilon \ge \gamma \ge \frac{\theta_b \cdot \capacity{\worker}}{\alpha_{\worker} }- \frac{1}{n\alpha}$$

Now, by solving for $\alpha_{\worker}$, we get:

$$\alpha_{\worker} \ge \frac{\theta _b \cdot \capacity{\worker}}{\frac{1}{n\alpha}+\epsilon}
\ge \theta_b\cdot\capacity{\worker}n\alpha (1 - \epsilon n \alpha),$$
where we use the Bernoulli's inequality $(1 + \epsilon n \alpha)^{-1} \ge (1 - \epsilon n \alpha)$ to obtain the above second inequality.
%\notes{not sure if it's clear how you get the final approximation}

Notice that the above inequality gives the lower bound on the number of virtual workers assigned to an overloaded
worker. Since its optimal number of virtual workers is $\alpha_{\worker}^* = \capacity{\worker}n\alpha$,
we can see that $\alpha_{\worker} / \alpha_{\worker}^* \ge \theta_b (1 - \epsilon n\alpha)$, which is close to $1$
since $\epsilon \ll 1/(\alpha n)$. This gives an interesting property that once we know
a worker is overloaded, we can be sure that its number of virtual workers is close to the optimal allocation. 
Thus, the sources can probe the capacity of workers by assigning virtual workers (taken from overloaded workers) to workers that have not reported becoming overloaded,
or if there is no such one, to those that reported becoming overloaded least recently.
Also notice that by letting $\theta_b = (1 + \epsilon n \alpha)$, we can guarantee that the overloaded workers are having at least
the optimal number of virtual workers they should have. However, when $\epsilon$ is large (due to bad load balancing mechanisms), or when $\alpha n$ is large (due to having many small virtual workers), $\theta_b > 1$ will become large. This will burden the overloaded workers
because they can only broadcast the overloaded cases when the threshold $\theta_b\cdot \capacity{\worker}$ is surpassed. This illustrates the tradeoff of load balancing mechanisms, with small imbalance factor $\epsilon$, and the right number of virtual workers (too many is not good) in our consistent grouping strategy.

\subsection{Memory with Consistent Grouping}

\kg generates the optimal memory footprint by forwarding each key to exactly one worker.
Similarly, \pkgs produces nearly optimal memory overhead by allowing at most two workers per key.
On the other end, \potc and \sg might assign each key to all the workers in the worst case, producing the memory footprint proportional to the number of workers.
Assume that $X_i$ is a random variable representing the minimum number of bins required for a ball $i$ with color $k_i$.
Further, assume a random variable $X$ representing the sum of number of bins required for the balls, i.e., $X=\sum\limits_{i=1}^{c} X_i$.
A trivial upper bounded for $X$ in case of shuffle grouping is given by: %for \sg equals $min(\ceil{p_i\cdot m},n)$.
\begin{equation} \label{eq2}
\begin{split}
\expect[X] & = \sum\limits_{i=1}^{c} min(\ceil{p_i\cdot m},n)
\end{split}
\end{equation} 

\porc allows a tradeoff between imbalance and memory using the parameter $\epsilon$.
To analyze the memory footprint of \porc, we answer a very simple question: \textit{What is the probability that a key is replicated on all the workers?}
For this to happen, the load of $n-1$ workers should exceed by $(1+\epsilon)$ of the average load.
Only then a key is replicated on all the workers.
However, for a sufficiently large value of $\epsilon$, i.e., $\epsilon > \frac{1}{n-1}$ this can not happen.
A trivial lower bound on the number of bins required for a ball $i$ with color $k_i$ is $\expect[X_i] = \ceil{\frac{p_i\cdot n}{(1+\epsilon)}}$.
Then,
\begin{equation} \label{eq2}
\begin{split}
\expect[X] & = \sum\limits_{i=1}^{c} \ceil{\frac{p_i\cdot n}{(1+\epsilon)}}
\end{split}
\end{equation}

%Assume that $Y_w$ be a random variable representing the load of a bin.
%We analyze the probability that given there are $f$ full bins, i.e., $f= \mid \{Y_w \colon Y_w > (1+\epsilon)\cdot \frac{m}{n}\}\mid$ among $n$ bins, the probability of a ball must try $k$ number of hash functions before finding a bin with spare capacity is:
%
%$$\prob(Z = k | F = f) = \Big(\frac{f}{n}\Big)^{k-1} \Big( 1 - \frac{f}{n}\Big)$$

This discussion provides the basic intuition on why the memory overhead of \porc is lower than \sg and \potc.
We plan to consider the detailed analysis in future work.

% !TEX root =  ../main.tex

\section{Evaluation}
\label{sec:eval}

%We conduct an extensive empirical evaluation of the proposed algorithm, and provide comparisons with the existing solutions.
We assess the performance of our proposal by using both simulations and a real deployment.
In so doing, we answer the following questions:
\begin{squishlist}
\item[\textbf{Q1:}]
What is a good set of values for the parameters of \cg?
%\notes{memory cost?computation cost?clarify}
\item[\textbf{Q2:}]
How does \cg perform compared to other schemes?
%\notes{what imbalance are we talking here? Consider rephrasing with respect to resource utilization?}
\item[\textbf{Q3:}]
How does \cg adapt to changes in input stream and resources?
\item[\textbf{Q4:}]
What is the overall effect of \cg on applications deployed on a real \dspe?
\end{squishlist}

\subsection{Experimental Setup}
\begin{table}[t]
\caption{Summary of the datasets used in experiments. Note: Percentage of messages having the most frequent key ($p_1$). %, arrival rate (\arrivalrate) for data streams. %It provides the number of edges, number of vertices, and maximum degree for the graphs.
} 
\centering
\small
\begin{tabular}{l c c c c }
\toprule
Stream		&	Symbol	&	Messages		&	Keys			& 	$p_1$	(\%) \\
\midrule
Wikipedia		&	WP		&	\num{22}M	&	\num{2,9}M	&	\num{9,32}  \\
Twitter		&	TW		&	\num{1,2}G	&	\num{31}M	&	\num{2,67}	 \\
%Cashtags		&	CT		&	\num{1,4}M	&	\num{2,9}k	&	\num{3,29}	& 338 \\
Zipf			&	ZF		&    \num{10}M   	&  \num{100}k &   $\propto$ $\frac{1} { \sum{x^{-z}} }$ \\
%\midrule
%Amazon		& AM		&  \num{925}k & \num{334}k	& -  \\
%DBLP		&  DB    &   \num{1}M & \num{317}k & -  \\
%Wiki Graph  & WG  & \num{28}M & \num{1,7}M & - \\
%Live Journal  & LJ & \num{34}M  &	\num{3,9}M & -\\
\bottomrule
\end{tabular}
\label{tab:summary-datasets}
\end{table}

\begin{figure}[t]
\begin{center}
\includegraphics[width=0.75\columnwidth]{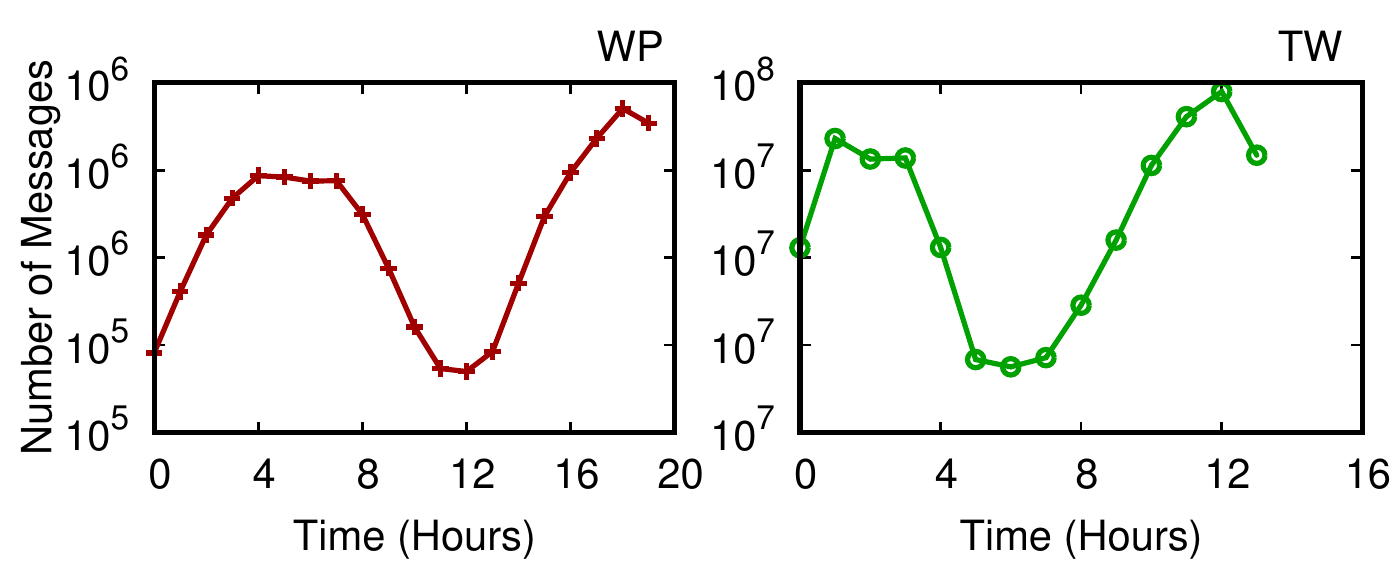}
\caption{Number of messages per hour for WP and TW datasets.
%\notes{what is number of events? Number of messages? And in caption you say number of requests... please be consistent}
\label{fig:data-rate}}
\end{center}
\vspace{-\baselineskip}
\end{figure}

\begin{table}[t]
\caption{Notation for the algorithms tested.}
%\vspace{-2mm}
\centering
\small
\begin{tabular}{l l }
\toprule
Symbol	&	Algorithm		\\
\midrule
%Hashing & H \\
\kg & Key Grouping	 \\
%\ch & Consistent Hashing	 \\
\pkgs & Partial Key Grouping		 \\
\potc   & Power of Two Choices   \\
\porc   & Power of Random Choices   \\
\ch   & Consistent Hashing with Bounded Load   \\
\sg   & Shuffle Grouping   \\
\cg   & Consistent Grouping  \\
\bottomrule
\end{tabular}
\label{tab:summary-algorithm}
\end{table}

\spara{Datasets.}
Table~\ref{tab:summary-datasets} summarizes the datasets used.
In particular, our goal is to be able to produce skewness in the input stream. % and 2) skewness in the processing time for each tuple.
We use two main real data streams, one from \emph{Wikipedia} and one from \emph{Twitter}.
These datasets were chosen for their large size, different degree of skewness, and different set of applications in Web and online social network domains.
The Wikipedia dataset (WP)\footnote{\url{http://www.wikibench.eu/?page\_id=60}} is a log of the pages visited during a day in January 2008. %~\citep{urdaneta09wikipedia}.
Each visit is a message and the page's URL represents its key.
The Twitter dataset (TW) is a sample of tweets crawled during July 2012.
We split each tweet into words, which are used as the key for the message.
%Figure \ref{fig:arrival-rate} shows the arrival rate for WP and TW dataset.
%\begin{figure}[t]
%\begin{center}
%\includegraphics[width=0.8\columnwidth]{arrival-rate}
%\caption{Plots showing the number of messages per minute for Wikipedia (WP)  and Twitter (TW) dataset.
%\label{fig:arrival-rate}}
%\end{center}
%\vspace{-\baselineskip}
%\end{figure}
%Additionally, we use a Twitter dataset that comprises of tweets crawled in November 2013.
%The keys for the messages are the \emph{cashtags} in these tweets.
%A \emph{cashtag} is a ticker symbol used in the stock market to identify a publicly traded company preceded by the dollar sign (e.g., \$AAPL for Apple).
%As shown in Figure~\ref{fig:key-frequency-distribution-tickers}, popular cash tags change from week to week.
%This dataset allows to study the effect of shift of skew in the key distribution.
%\begin{figure}[t]
%\begin{center}
%	\includegraphics[width=\columnwidth]{img/variation_of_toptickers.pdf}
%	\caption{Frequency of tweets for the top 5 tickers in CT.
%	The most frequent keys change throughout time.}
%	\label{fig:key-frequency-distribution-tickers}
%\end{center}
%\vspace{-\baselineskip}
%\end{figure}
%To mimic the skew in the processing time for each tuple, we use the two hop count query (discussed in the Section \ref{sec: challenges}).
%We use several publicly available graph datasets to construct the graph and execute the two-hop neighborhood count query on top of this graph (see Table~\ref{tab:summary-datasets} for description). 
Figure \ref{fig:data-rate} reports the ingestion rate of the streams in terms of number of messages per hour.
Lastly, we generate synthetic datasets with keys following Zipf distributions with exponent in the range $z = \{ 0.1, \ldots, 2.0 \}$ with 100$k$ unique keys.
%\notes{here you say different number of unique keys, but you say only 100k, and then in the table you show varying number. Please check and be consistent}

%We combine the zipf dataset along with both data streams and graphs to enable executing the complete \dagr of an application.
%Each unique key of rank \emph{r} appears with frequency $f$ as follows:
%$$f(r,K,z) = \frac {1/r^z }{\sum_{x=1}^{K} (1/x^z)}.$$
%Finally, we experiment on two additional datasets comprised of directed graphs\footnote{\url{http://snap.stanford.edu/data/}} (WG, LJ). 
%We use the edges in the graph as messages and the vertices as keys.

\spara{Simulation and Real Deployment.}
We process the datasets by simulating the \dagr presented in Figure~\ref{fig:vnb}.
The stream is composed of timestamped keys that are read by multiple independent sources (\sources) via shuffle grouping, unless otherwise specified.
The sources forward the received keys to the workers (\workers) downstream.
In our simulations we assume that the sources perform data extraction and transformation, while the workers perform data aggregation, which is the most computationally expensive part of the \dagr.
Thus, the workers are the bottleneck in the \dagr and the focus for the load balancing.
Note that for simulation, we ignore the network latency.
The selected workloads represent a variety of streaming applications. 
In particular, any application that performs reduce-by-key or group-by operation follows a similar pattern.

\spara{Algorithms.}
Table~\ref{tab:summary-algorithm} defines the notations used for the different algorithms.
We use a $64$-bit Murmur hash function for implementation of \kg to minimize the probability of collisions.
Unlike the algorithms in Table~\ref{tab:summary-algorithm}, other related load balancing algorithms~\citep{shah2003flux,cherniack2003scalable,xing2005dynamic,balkesen2013adaptive,castro2013integrating} require the \dspe to support operator migration.
Many top \dspes, such as Apache Storm, do not support migration.
Thus, we omit these algorithms from the evaluation.

\spara{Metrics.}
Table \ref{tab:algorithmic-metric} defines the metrics used for the evaluation of the performance of different algorithms.

\begin{table}[t]
\caption{Metric used for evaluation of the algorithms.}
\label{tab:algorithmic-metric}
%\vspace{-2mm}
\centering
\small
\begin{tabular}{l l }
\toprule
Metric		& Description \\
\midrule
Memory Cost & Replication cost of the keys \\
Queue Length & Number of messages in the queue \\
Resource Utilization & Ratio between number of messages \\
& and capacity of worker.\\
Imbalance &  Difference between the maximum  and the \\
& average resource utilization.\\
\midrule
Execution Latency &  Difference between arrival and finish time. \\ 
Throughput &  Number of messages processed per second.\\
\bottomrule
\end{tabular}
\end{table}

\spara{Monitoring Performance.}
For \cg, each worker requires monitoring its resource utilization that enables the fair message assignment.
In case of simulations, we define the resource utilization as the ratio between the number of assigned messages and the capacity of a worker.
We define the notion of idle and busy worker using the $\utilization{\worker}{t} < 0.75\cdot\capacity{\worker}$ and $ \utilization{\worker}{t} > 0.85\cdot\capacity{\worker}$ thresholds respectively.
For the real-world experiments, we suggest using the queue length as a parameter for monitoring the resource utilization. 
In particular, the resource utilization is defined by: 
$$ \utilization{\worker}{t} = \dfrac{\#\text{tuples in the queue}}{\text{input queue capacity}} = \dfrac{\load{\worker}{t}}{\capacity{w}}$$

The choice of the parameter was motivated by its availability in the standard Apache Storm distribution (ver 1.0.2).

%\notes{You say here resource utilization, then in the plots you say memory overhead or imbalance.. Also in text you say resource utilization. Please be consistent}
%
%\notes{I would like to see clearly defined experimental metrics measured here}

%\notes{I would also describe here (or earlier when we present the experimental questions) in a small paragraph the methodology we follow, the general "flow" of the experiments. I.e., we first perform experiments to tune our parameters by considering key mapping to virtual workers, then we choose the best method, then we perform experiments to find the best method to map virtual workers to workers, then we do experiments with varying stream skew and heterogeneity, and finally we apply our lessons on a real DSPE}

%\spara{Experimental Methodology.}
%We first perform experiments to tune our parameters by considering key mapping to virtual workers.
%Afterwards, we choose the best method and perform experiments to find the best method to map virtual workers to workers, then we do experiments with varying stream skew and heterogeneity, and finally we apply our lessons on a real DSPE
\subsection{Experimental Results}

\begin{figure}[t]
\begin{center}
\includegraphics[width=0.85\columnwidth]{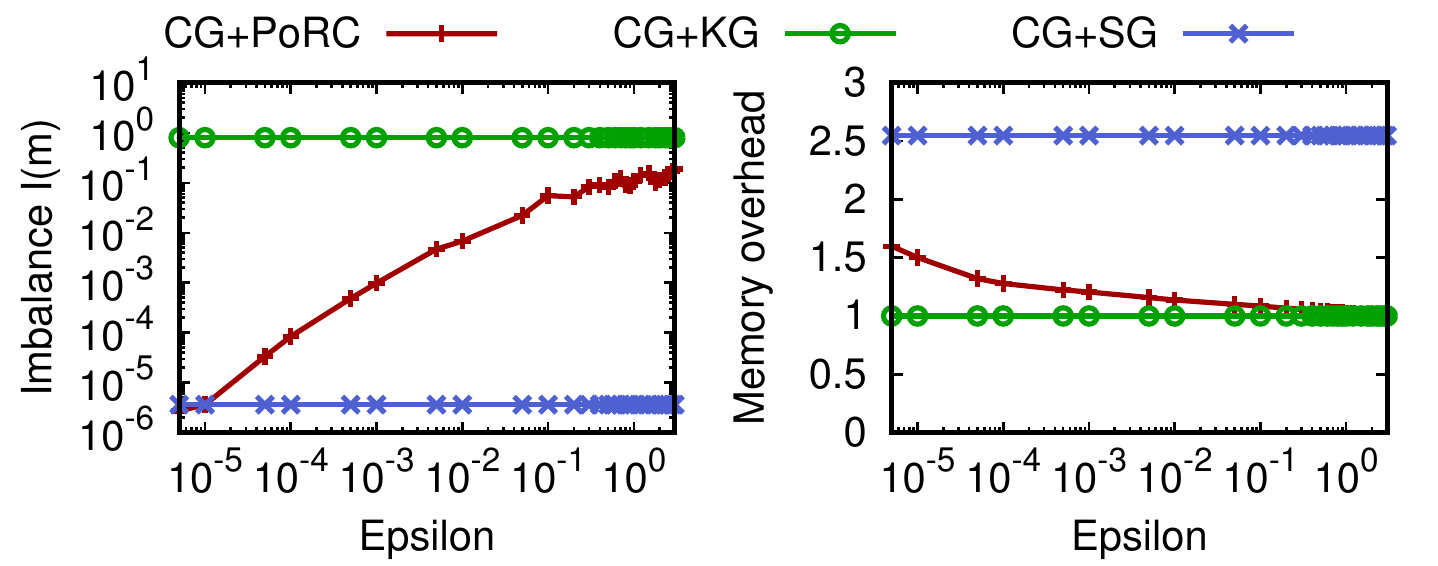}
\caption{Experiment reporting the imbalance and the memory overhead for different values of epsilon. 
The setup includes $10$ workers, each having $10$ virtual workers mapped to it.
\label{fig:epsilon}}
\end{center}
\vspace{-\baselineskip}
\end{figure}

\textbf{Q1:} 
In the first experiment, we simulate the \cg scheme by varying the value of $\epsilon$ and fixing the number of sources to $1$ and the number of workers to $10$.
Each worker is homogeneous and the number of virtual workers per worker are set to $10$. 
We select the WP dataset and simulate \cg for different values of $\epsilon$. %, i.e., $\epsilon \in \{0.001,\ldots, 3.0\}$.
We leverage \kg, \sg and \porc for task assignment to the virtual workers.
Figure \ref{fig:epsilon} reports the imbalance and the memory overhead for the experiment.
The results verify our claim that epsilon provides a trade-off between imbalance and memory. 
In particular, \cg generates low imbalance at lower values of epsilon and produces low memory footprint for higher values of epsilon.
Also, the experiment shows that \cg is able to interpolate well between the \kg and \sg schemes.
Based on this experiment, we use the value of $\epsilon$=$0.01$ henceforth as it provides a middle ground between memory and imbalance.
%\notes{I am not clear why you pick this value. You want some middle ground between memory overhead and imbalance? What about the fact that we didn't test heterogeneity yet?}

\begin{figure}[t]
\begin{center}
\includegraphics[width=0.85\columnwidth]{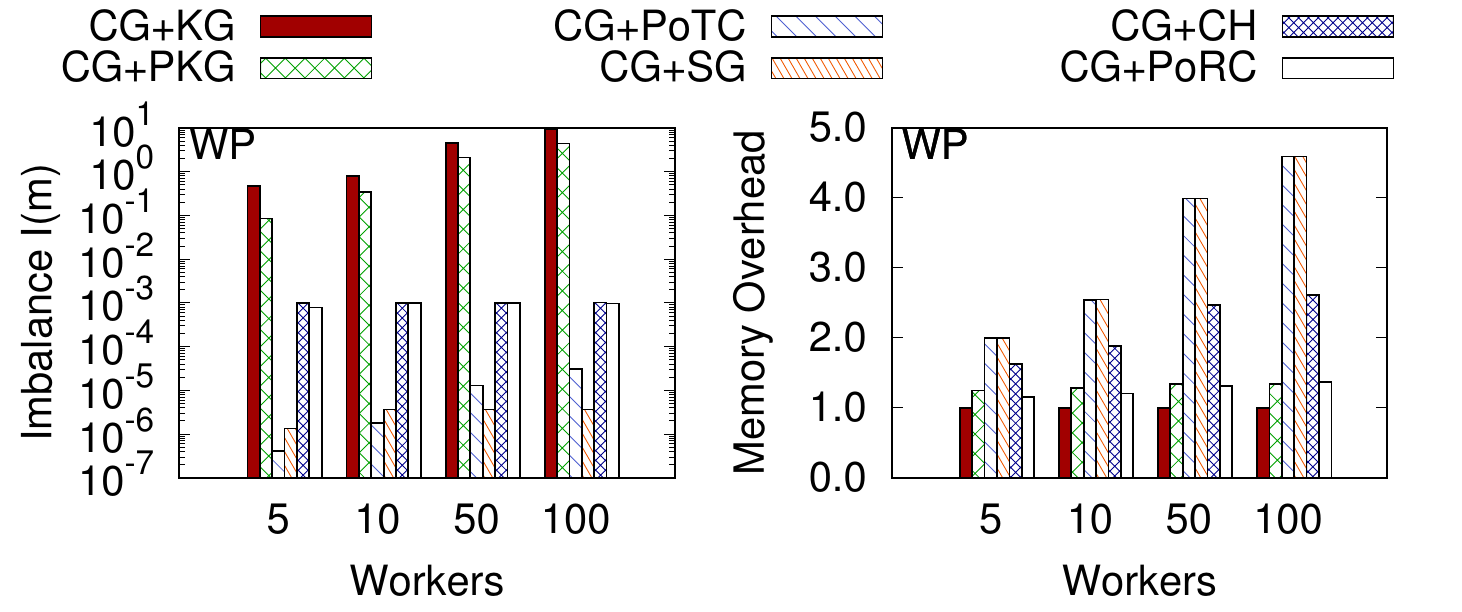}
\caption{Normalized imbalance and memory overhead comparing several assignment strategies along with consistent grouping on a homogeneous cluster with $5$, $10$, $50$ and $100$ workers using WP dataset. 
Each worker spawns $10$ virtual workers and the $\epsilon=0.01$ for \ch and \porc.
%\notes{It is useful to say here what you investigate with 3 words}
\label{fig:load-imbalance-cg-bmf}}
\end{center}
\vspace{-\baselineskip}
\end{figure}

Next, we analyze the allocation strategies, i.e., \kg, \pkgs, \potc, \porc, \ch and \sg. 
We simulate an experiment on a homogeneous cluster  with $5$, $10$, $50$ and $100$ workers using the WP dataset.
The number of virtual workers per worker are set to $10$, i.e., equivalent to splitting the keys into $50$, $100$, $500$ and $1000$ bins. 
For \ch and \porc, we set $\epsilon=0.01$.
Figure \ref{fig:load-imbalance-cg-bmf} shows the imbalance after the assignment of the streams.
Results show that \kg and \pkgs generate high imbalance, whereas \potc and \sg generate nearly perfect load balance. 
Both \ch and \porc bound the imbalance close to a constant factor from the value of $\epsilon$.
The imbalance in case of \kg and \pkgs grows linearly with the increase in the number of workers. %skewness and the number of workers.
This behavior is due to the fact that both these schemes restrict a single key to a constant number of workers.
\ch and \porc bound the imbalance upto a constant factor for each bin.
\potc and \sg achieve near perfect imbalance by exploiting all the possible workers.
Interestingly, \porc achieves bounded imbalance while keeping the memory footprint as low as \pkgs, as shown in Figure \ref{fig:load-imbalance-cg-bmf}.
In particular, \porc generates nearly perfect memory footprint and operates very close to \kg.
The gain in memory footprint depends on the distribution of the workload and the size of the deployment, and achieving gains in orders of magnitude is not always possible.
Henceforth, we leverage \porc for the next experiments and analyze consistent grouping.
%\notes{It makes me worried that we didn't test heterogeneity of workers up to here, but we have already fixed epsilon and the type of mapping to PORC}
%\notes{Also, it makes me worried that we don't really have an adaptive way to assign bins (or virtual workers) to actual workers}

\begin{figure}[t]
\begin{center}
\includegraphics[width=0.85\columnwidth]{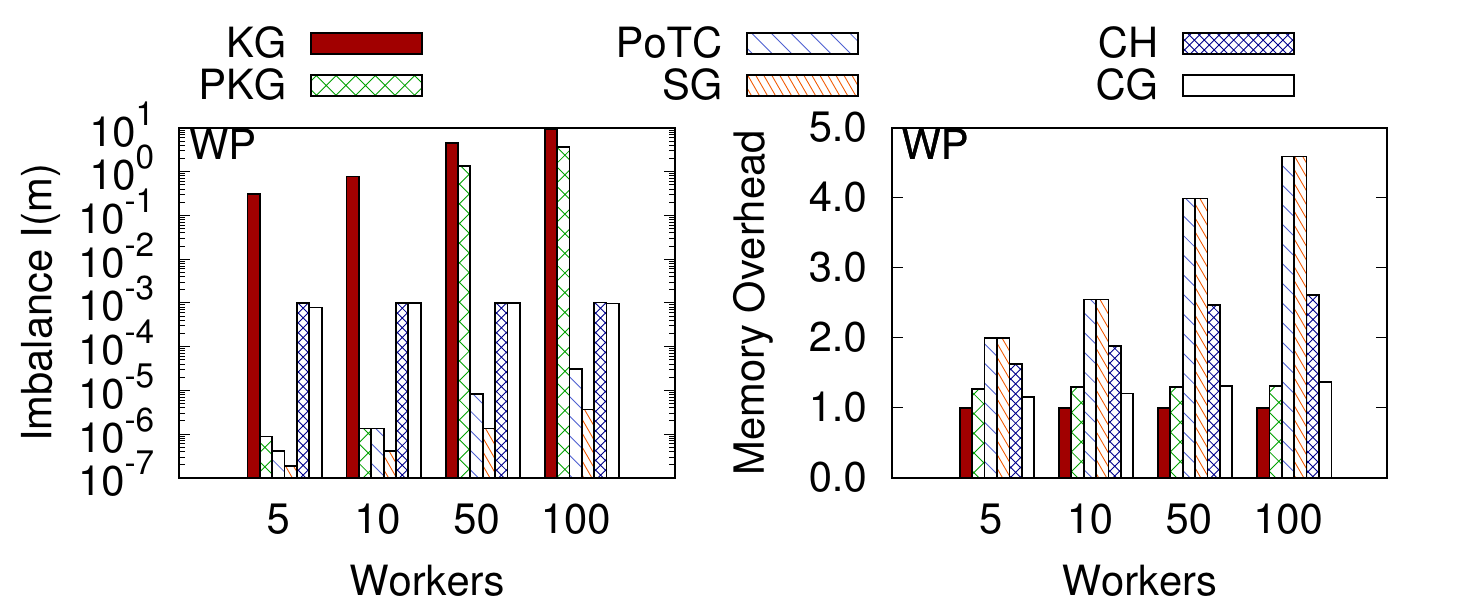}
\caption{Normalized imbalance and memory overhead comparing different grouping strategies on a homogeneous cluster with $5$, $10$, $50$ and $100$ workers using the WP dataset. 
Each worker spawns $10$ virtual workers and the $\epsilon=0.01$ for \ch and \porc.
%\notes{It is useful to say here what you investigate with 3 words}
%\notes{Right now this caption reads exactly like Figure 7}
\label{fig:load-imbalance-homo}}
\end{center}
\vspace{-\baselineskip}
\end{figure}

\textbf{Q2:}
To answer this question, we compare the imbalance and the memory overhead of \cg with \kg, \pkgs, \potc, \ch and \sg. 
We simulate the DAG using the WP dataset and report the value of imbalance measured at the end of the simulation. % 
The cluster consists of different number of workers, i.e., $5$, $10$, $50$ and $100$ workers.
Each experiment considers a cluster of homogeneous machines.
For \cg and \ch, we set the value of epsilon equal to $0.01$.
Figure \ref{fig:load-imbalance-homo} reports the imbalance and the memory overhead for different schemes (note the log scale).
%\notes{looks out of order figure}
Results show that \kg performs the worst in terms of the imbalance while generating the optimal memory footprint.
\pkgs on the other hand provides nearly perfect imbalance and optimal memory footprint for smaller deployments, i.e., $5$ and $10$ workers.
However, the imbalance grows as the number of workers increase. 
\potc and \sg provide very similar performance, i.e., provide nearly perfect imbalance and generate higher memory footprint.
\ch provides bounded imbalance and reduces the memory footprint compared to \potc and \sg.
\cg provides the bounded imbalance and improves the memory footprint compared to \ch.
This behavior is due to the fact that \cg leverages randomness to redistribute the messages once the principal worker reaches the capacity, whereas \ch always choose the next worker in the ring.

\begin{figure}[t]
\begin{center}
\includegraphics[width=\columnwidth]{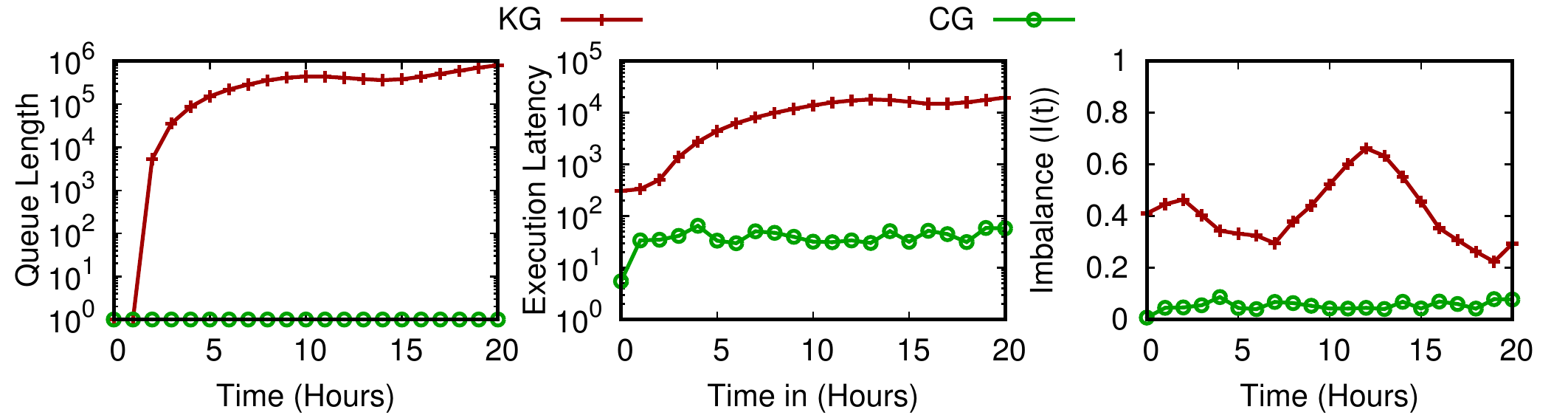}
\caption{Effect on queue length, execution latency and resource utilization on a homogeneous cluster with 10 workers for \kg and \cg using the WP dataset.
Each worker spawns $10$ virtual workers and the $\epsilon$ equals $0.01$ for \cg.
%\notes{check y-label for execution latency}
\label{fig:input-queue-homo}}
\end{center}
\vspace{-\baselineskip}
\end{figure}

Additionally, we report the queue length, execution latency and the resource utilization among the workers by setting the capacity of the workers in a way that each worker operates at $80\%$ of the capacity using shuffle grouping.
We report each metric as a difference between the maximum and minimum value.
Note the the difference between the maximum and minimum resource utilization represents the imbalance.
Due to space restriction, we only report the results for 10 workers.
For comparison, we also simulate and report \kg and \cg.
Note that as \pkgs, \potc and \sg provide nearly perfect load balance, we do not report their results (the different between maximum and average queue length, execution latency and resource utilization equals 0).
%\notes{is that fair? I don't get it}
We simulate the WP dataset, set the value of $\epsilon$ equal to $0.01$ and set the number of virtual workers per worker equal to $10$ for \cg.
Figure \ref{fig:input-queue-homo} shows the results of the experiment over time.
Results show that the difference between the maximum and minimum queue length and execution latency increases over time using \kg, whereas \cg keeps both queue length and execution latency very low.
Also, the imbalance is high for \kg, whereas \cg keeps the imbalance to close to zero.

%\notes{see previous comment on resource utilization vs. memory overhed vs. imbalance vs. execution latency. Please define them clearly in the beginning of the section and use consistent labels in plots and discussion in the text. Otherwise the reviewers will be confused and reject it}

\begin{figure}[t]
\begin{center}
\includegraphics[width=\columnwidth]{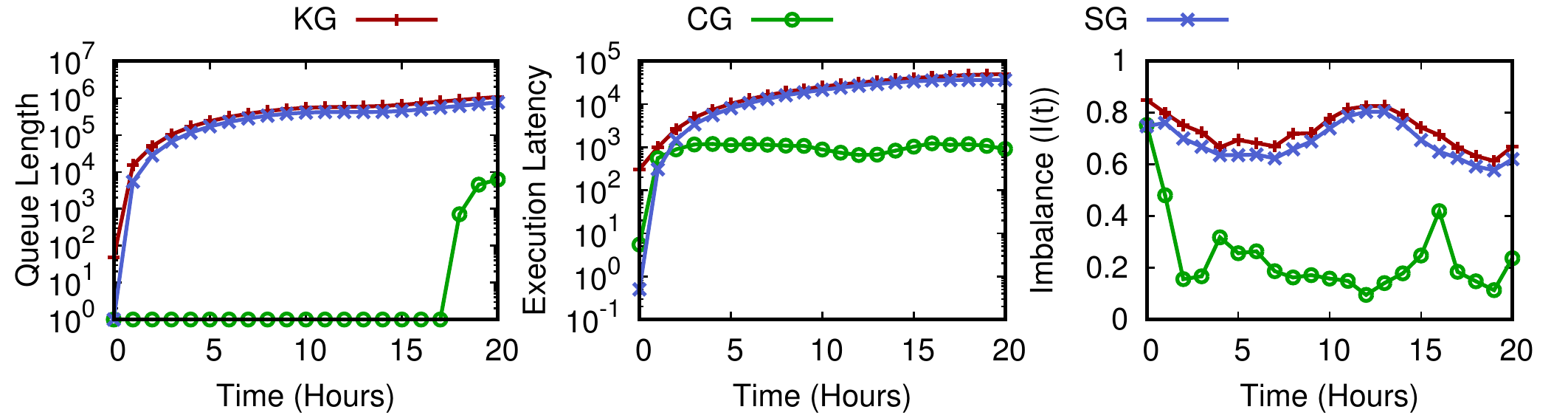}
\caption{Effect on queue length, execution latency and resource utilization due to heterogeneity in the cluster for \kg, \cg and \sg.
%\notes{check the y-label, execution latency}
\label{fig:input-queue-hetero}}
\end{center}
\vspace{-\baselineskip}
\end{figure}

Next, we mimic the heterogeneity in the cluster by assuming a cluster consisting of $n$ machines in which $y$ machines are $z$ times more powerful than rest of the machines. 
In particular, we vary the value of $z$ between $2$ to $10$ and vary the value of $y$ between $1$ to $n-1$.
For instance, when $y=1$ and $z=2$, a machine in a $10$-machine cluster has twice the capacity than all the other nine machines.
We simulate the \kg, \sg, \cg for comparison and use the value of epsilon equal to $0.01$.
In case of \cg, each worker is initialized with $10$ virtual workers. 
We observe similar behavior in all the configurations and report only a single iteration with $y=3$ and $z=5$.
Figure \ref{fig:input-queue-hetero} reports the queue length, execution latency and resource utilization for the three approaches.
Results show that queue length and execution latency grow for \kg and \sg.
Similarly, the imbalance is pretty high for these approaches.
On the other hand, \cg provides the lower queue length and execution latency.
Also, it keeps the imbalance close to zero. % results by keeping the queue length, execution latency and resource utilization low.
Note that there is a spike after $17$ hours for queue length, which is due to the fact that we leverage the resource utilization as a metric to segregate between idle and busy workers. %as a metric for the migration of virtual workers.

%\notes{see previous comment on resource utilization vs. memory overhed vs. imbalance vs. execution latency. Please define them clearly in the beginning of the section and use consistent labels in plots and discussion in the text. Otherwise the reviewers will be confused and reject it}

\begin{figure}[t]
\begin{center}
\includegraphics[width=0.85\columnwidth]{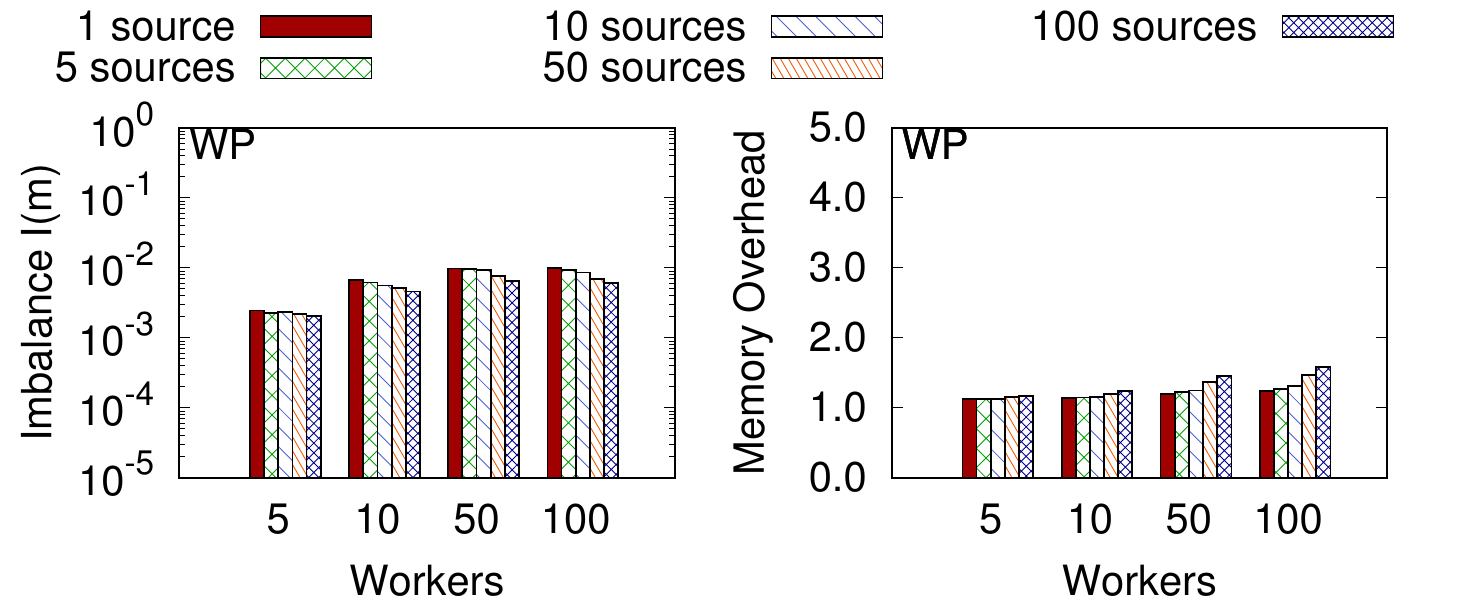}
\caption{Normalized imbalance and memory overhead on a homogeneous cluster with $5$, $10$, $50$ and $100$ workers with using $1$, $10$, $50$, and $100$ sources using WP dataset. 
Each worker spawns $10$ virtual workers and the $\epsilon$ equals $0.01$ for \cg.
\label{fig:load-imbalance-sources}}
\end{center}
\vspace{-\baselineskip}
\end{figure}

\textbf{Q3:} 
Further, we evaluate the performance of \cg by increasing the number of sources.
In particular, we compare the performance of different deployments using $1$, $10$, $50$, and $100$ sources.
For assignment of messages to sources, we use \sg.
Figure \ref{fig:load-imbalance-sources} reports the performance of \cg in terms of imbalance and memory overhead.
Results show that both imbalance and memory footprint almost remain the same on a log scale by both increasing the number of workers and number of sources.
Therefore, we can conclude that \cg is able to provide similar performance even under higher number of sources and workers.

\begin{figure}[t]
\begin{center}
\includegraphics[width=\columnwidth]{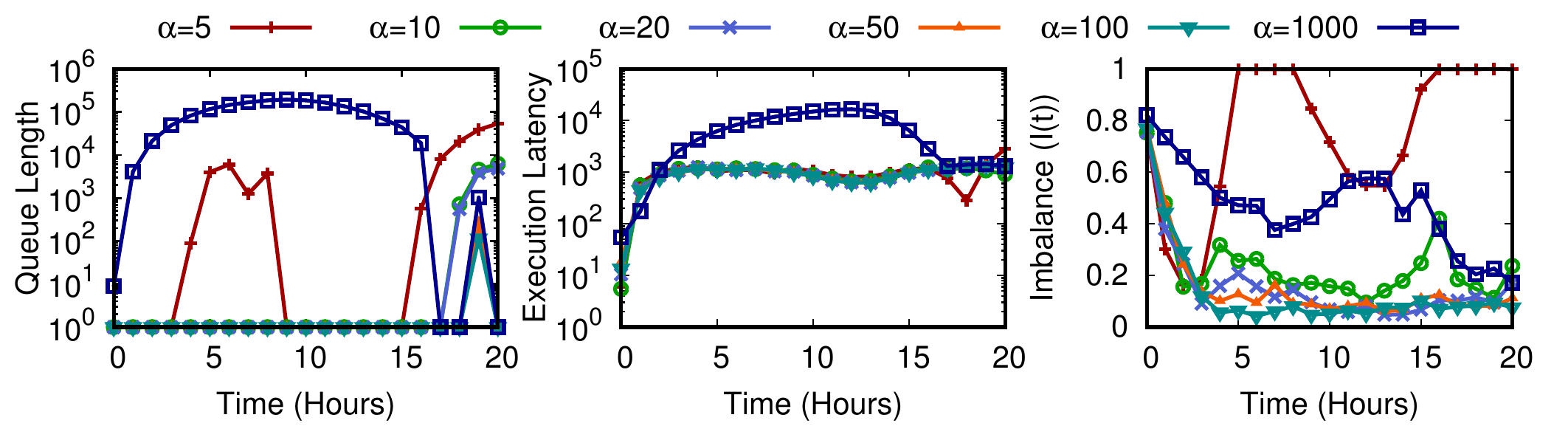}
\caption{Queue length, execution latency and resource utilization for different number of virtual workers.
\label{fig:virtual-worker}}
\end{center}
\vspace{-\baselineskip}
\end{figure}

Further, we study the behavior of \cg on the number of virtual workers.
We reuse the configuration for the experiment reported in Figure \ref{fig:input-queue-hetero} and report the queue length, execution latency and resource utilization for \cg, i.e., $y=3$ and $z=5$.
We perform the experiment using number of virtual workers equal to $5$, $10$, $20$, $50$, $100$ and $1000$.
Figure \ref{fig:virtual-worker} shows that setting the number of virtual workers to a value of $5$ does not provide desired results. 
This is due to the fact that there are not enough virtual workers to move around the workers.
Similarly, when the number of virtual workers are equal to $1000$, the system takes longer time to converge, hence impacting the performance.
Executions using $10$ and $20$ virtual workers provide similar performance.
Lastly, the execution using $100$ virtual workers generate the best results.

\begin{figure}[t]
\begin{center}
\includegraphics[width=\columnwidth]{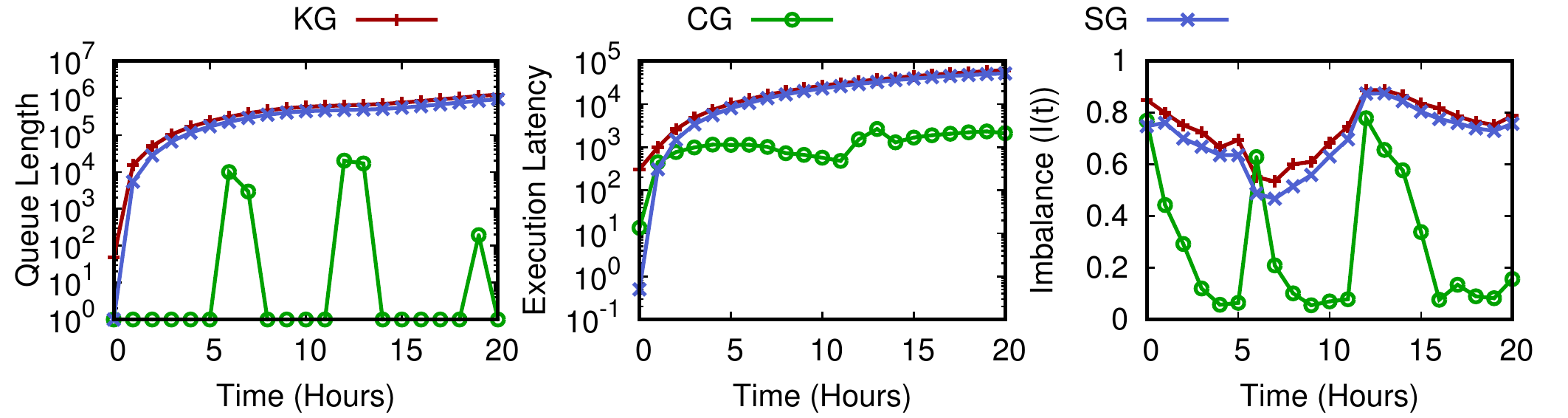}
\caption{Queue length, execution latency and resource utilization for when resources are changing over time.
The resources change after processing \num{6}M and \num{12}M messages.
\label{fig:dynamic-resources}}
\end{center}
\vspace{-\baselineskip}
\end{figure}

Next, we study the performance of \cg by dynamically changing the resources over time.
To initialize the resources, we reuse the configuration from the previous experiment and change the capacity of resources twice during the execution, i.e., after processing \num{6}M and \num{12}M messages.
Concretely, we change the values of $y$ and $z$ (represented as \{$y$, $z$\}) after \num{6}M and \num{12}M messages to \{$5,4$\} and \{$2,10$\} respectively.
We execute the experiment for $100$ virtual workers and change the resources in a way that the sum of resources remains the same.
Also, we report the results of \kg and \sg for comparison.
Figure \ref{fig:dynamic-resources} reports the queue length, execution latency and resource utilization of the experiment.
Results show that \cg adapts very efficiently to the change in resources.

%\notes{see previous comment on resource utilization vs. memory overhed vs. imbalance vs. execution latency. Please define them clearly in the beginning of the section and use consistent labels in plots and discussion in the text. Otherwise the reviewers will be confused and reject it}

\textbf{Q4:}
Lastly, we study the effect of \cg on streaming applications deployed on an Apache Storm cluster running in a private cloud.
The storm cluster consists of $8$ medium sized machines with $2$ virtual CPUs and $4$ GB of memory each. %, which runs the storm cluster.
Moreover, a Kafka cluster with $8$ partitions is used as a data source.
We perform experiments to compare \cg, \pkgs, \kg, and \sg on the TW dataset.
The parameters are selected in a way that the number of sources and workers match the number of executors in the Storm cluster.
In this experiment, we use a topology configuration with $8$ sources and $24$ workers.
We report overall throughput, end-to-end latency and memory footprint.

\begin{figure}[t]
\begin{center}
\includegraphics[width=\columnwidth]{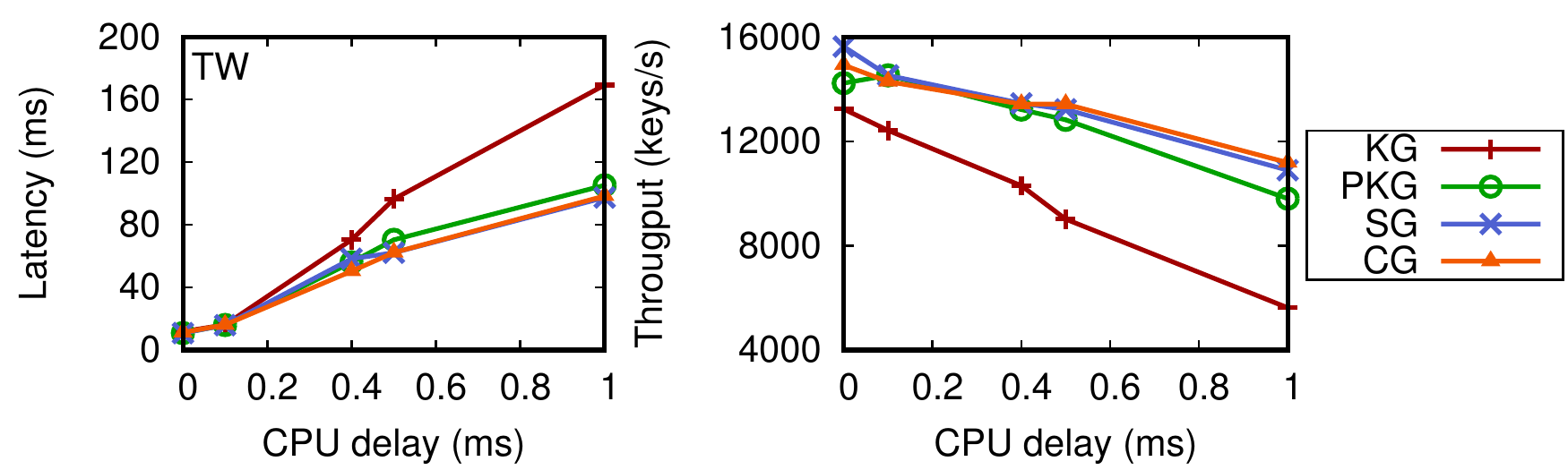}
\caption{Throughput and latency for TW dataset on a homogenous Storm cluster.
\label{fig:wp-cluster-homo}}
\end{center}
\vspace{-\baselineskip}
\end{figure}

\begin{figure}[t]
\begin{center}
\includegraphics[width=\columnwidth]{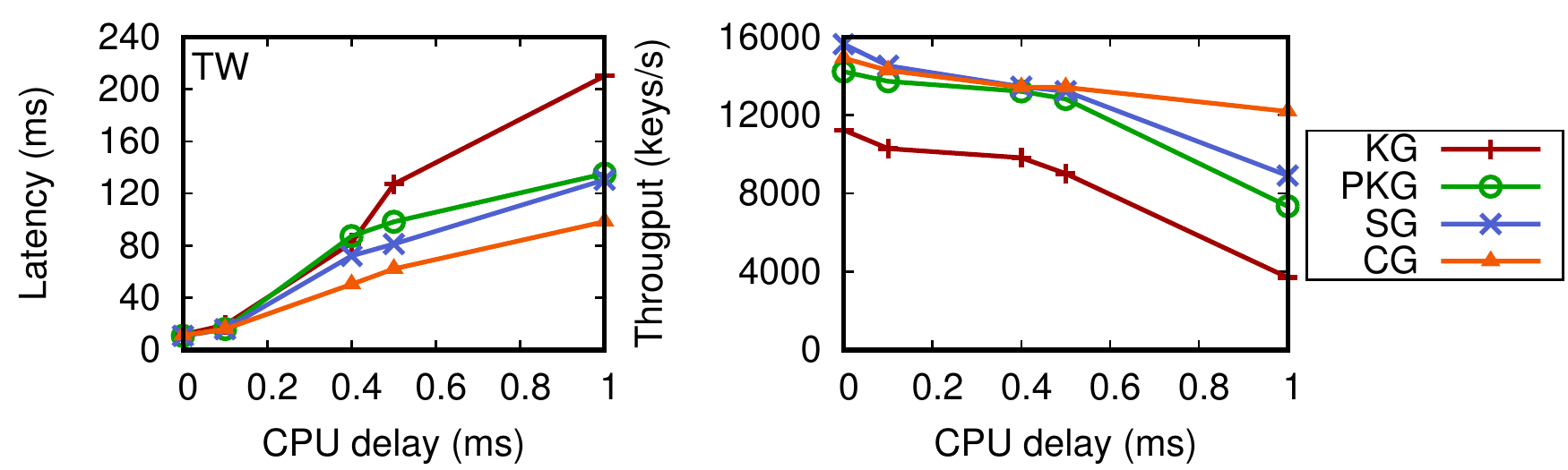}
\caption{Throughput and latency for TW dataset on a heterogenous Storm cluster.
\label{fig:wp-cluster-hetero}}
\end{center}
\vspace{-\baselineskip}
\end{figure}

In the first experiment, we evaluate the performance of the algorithms in a homogeneous cluster. 
We emulate different levels of CPU consumption per key, by adding a fixed delay to the processing.
We prefer this solution over implementing a specific application to control better the load on the workers.
We choose a range that can bring our configuration to a saturation point, although the raw numbers would vary for different setups.
Even though real deployments rarely operate at saturation point, \cg allows better resource utilization, therefore supporting the same workload on a smaller number of machines, but working on a higher overall load point each.
In this case, the minimum delay ($0.1$ms) corresponds approximately to reading 400kB sequentially from memory, while the maximum delay ($1$ms) to $\frac{1}{10}$-th of a disk seek.\footnote{\url{http://brenocon.com/dean_perf.html}}
Nevertheless, even more expensive tasks exist: parsing a sentence with NLP tools can take up to $500$ms.\footnote{\url{http://nlp.stanford.edu/software/parser-faq.shtml\#n}}

Figure \ref{fig:wp-cluster-homo} reports the throughput and end-to-end latency for the TW dataset on the homogenous cluster.
Also, during the experiment, \kg was consuming $7\%$ of memory in the cluster vs. $8.5\%$ for \pkgs and \cg, and $14\%$ for \sg.
Results shows that \kg provides low memory overhead but coupled with low throughput and high execution latency.
Alternatively, \pkgs, \sg and \cg provide superior performance in terms of throughput, latency and memory consumption.

Further, we evaluate the performance of \cg in the presence of heterogeneity in the cluster.
We use the {\bf cpulimit} application to change the resource capacity over time and monitor the behavior of different approaches in terms of throughput and end-to-end latency.
In particular, we limit the cpu resources of two of the executors to $30\%$ of the available CPU resources to mimic the heterogeneity in the cluster.
During the experiment, we give the system $10$ minutes grace period to reach a stable state before collecting the statistics.
Figure \ref{fig:wp-cluster-hetero} reports the throughput and the end-to-end latency of the experiment.
Results show that \cg outperforms the other approaches both in terms of throughput and end-to-end latency.
In particular, and compared to \kg, it provides up to $2\times$ better end-to-end latency and $3.44\times$ better performance in terms of throughput.

Overall, we observe that \cg is a very competitive solution with respect to \kg, \pkgs and \sg, performing much better with respect to throughput and end-to-end latency and imposing a small memory footprint, while at the same time tackling the problem of heterogeneity of available resources at the workers in the cluster.

%\notes{Anis, please fix the numbers we report in Intro and abstract}

%\notes{see previous comment on resource utilization vs. memory overhed vs. imbalance vs. execution latency. Please define them clearly in the beginning of the section and use consistent labels in plots and discussion in the text. Otherwise the reviewers will be confused and reject it}

%\input{sections/applications}
\section{Conclusion}

We studied the load balancing problem for streaming engines running in a heterogeneous cluster and processing varying workload. 
We proposed a novel partitioning strategy called {\emph Consistent Grouping}. % that adapts to traditional consistent hashing. 
\cg leverages two very simple, but extremely powerful ideas: \emph{power of random choices} and \emph{fair virtual worker assignment}.
It efficiently achieves fair load balancing for streaming applications processing skewed workloads. % and running on a heterogeneous cluster.
We provided a theoretical analysis of the proposed algorithm and showed via extensive empirical evaluation that the cg outperforms the state-of-the-art approaches. % with several orders of magnitude. % showing its effectiveness and applicability.
In particular, \cg achieves $3.44$x better performance in terms of latency compared to key grouping.

%\section{Acknowledgment}
%This paper is produced during the internship of the first author at IBM Research Tokyo.
%We would like to thank Scott Schneider, Tatsuhiro Chiba and Takeshi Yoshimura for their constructive feedback on the paper.

\balance
\bibliographystyle{abbrvnat}
\bibliography{biblio}

\begin{thebibliography}{40}
\providecommand{\natexlab}[1]{#1}
\providecommand{\url}[1]{\texttt{#1}}
\expandafter\ifx\csname urlstyle\endcsname\relax
  \providecommand{\doi}[1]{doi: #1}\else
  \providecommand{\doi}{doi: \begingroup \urlstyle{rm}\Url}\fi

\bibitem[Ahmad et~al.(2012)Ahmad, Chakradhar, Raghunathan, and
  Vijaykumar]{ahmad2012tarazu}
F.~Ahmad, S.~T. Chakradhar, A.~Raghunathan, and T.~Vijaykumar.
\newblock Tarazu: optimizing mapreduce on heterogeneous clusters.
\newblock In \emph{SIGARCH}, volume~40, pages 61--74. ACM, 2012.

\bibitem[Azar et~al.(1999)Azar, Broder, Karlin, and
  Upfal]{azar1999balanced-allocations}
Y.~Azar, A.~Z. Broder, A.~R. Karlin, and E.~Upfal.
\newblock Balanced allocations.
\newblock \emph{SIAM J. Comput.}, 29\penalty0 (1):\penalty0 180--200, 1999.

\bibitem[Balkesen et~al.(2013)Balkesen, Tatbul, and
  {\"O}zsu]{balkesen2013adaptive}
C.~Balkesen, N.~Tatbul, and M.~T. {\"O}zsu.
\newblock Adaptive input admission and management for parallel stream
  processing.
\newblock In \emph{DEBS}, pages 15--26. ACM, 2013.

\bibitem[Ben-Haim and Tom-Tov(2010)]{ben-haim2010spdt}
Y.~Ben-Haim and E.~Tom-Tov.
\newblock {A Streaming Parallel Decision Tree Algorithm}.
\newblock \emph{JMLR}, 11:\penalty0 849--872, 2010.

\bibitem[Berinde et~al.(2010)Berinde, Indyk, Cormode, and
  Strauss]{berinde2010heavyhitters}
R.~Berinde, P.~Indyk, G.~Cormode, and M.~J. Strauss.
\newblock {Space-optimal heavy hitters with strong error bounds}.
\newblock \emph{ACM Trans. Database Syst.}, 35\penalty0 (4):\penalty0 1--28,
  2010.

\bibitem[Cardellini et~al.(1999)Cardellini, Colajanni, and
  Philip]{cardellini1999dynamic}
V.~Cardellini, M.~Colajanni, and S.~Y. Philip.
\newblock Dynamic load balancing on web-server systems.
\newblock \emph{IEEE Internet computing}, 3\penalty0 (3):\penalty0 28, 1999.

\bibitem[Castro~Fernandez et~al.(2013)Castro~Fernandez, Migliavacca,
  Kalyvianaki, and Pietzuch]{castro2013integrating}
R.~Castro~Fernandez, M.~Migliavacca, E.~Kalyvianaki, and P.~Pietzuch.
\newblock Integrating scale out and fault tolerance in stream processing using
  operator state management.
\newblock In \emph{SIGMOD}, pages 725--736. ACM, 2013.

\bibitem[Chen et~al.(2010)Chen, Villa, Krishnamoorthy, and
  Gao]{chen2010dynamic}
L.~Chen, O.~Villa, S.~Krishnamoorthy, and G.~R. Gao.
\newblock Dynamic load balancing on single-and multi-gpu systems.
\newblock In \emph{IPDPS}, pages 1--12. IEEE, 2010.

\bibitem[Cherniack et~al.(2003)Cherniack, Balakrishnan, Balazinska, Carney,
  Cetintemel, Xing, and Zdonik]{cherniack2003scalable}
M.~Cherniack, H.~Balakrishnan, M.~Balazinska, D.~Carney, U.~Cetintemel,
  Y.~Xing, and S.~B. Zdonik.
\newblock Scalable distributed stream processing.
\newblock In \emph{CIDR}, volume~3, pages 257--268, 2003.

\bibitem[Das et~al.(2014)Das, Zhong, Stoica, and Shenker]{das2014adaptive}
T.~Das, Y.~Zhong, I.~Stoica, and S.~Shenker.
\newblock Adaptive stream processing using dynamic batch sizing.
\newblock In \emph{Proceedings of the ACM Symposium on Cloud Computing}, pages
  1--13. ACM, 2014.

\bibitem[Gedik(2014)]{gedik2014partitioning}
B.~Gedik.
\newblock Partitioning functions for stateful data parallelism in stream
  processing.
\newblock \emph{VLDB Journal}, 23\penalty0 (4):\penalty0 517--539, 2014.

\bibitem[Godfrey et~al.(2004)Godfrey, Lakshminarayanan, Surana, Karp, and
  Stoica]{godfrey2004load}
B.~Godfrey, K.~Lakshminarayanan, S.~Surana, R.~Karp, and I.~Stoica.
\newblock Load balancing in dynamic structured p2p systems.
\newblock In \emph{INFOCOM}, volume~4, pages 2253--2262. IEEE, 2004.

\bibitem[Godfrey and Stoica(2005)]{godfrey2005heterogeneity}
P.~B. Godfrey and I.~Stoica.
\newblock Heterogeneity and load balance in distributed hash tables.
\newblock In \emph{INFOCOM}, volume~1, pages 596--606. IEEE, 2005.

\bibitem[Gonzalez et~al.(2012)Gonzalez, Low, Gu, Bickson, and
  Guestrin]{gonzalez2012powergraph}
J.~E. Gonzalez, Y.~Low, H.~Gu, D.~Bickson, and C.~Guestrin.
\newblock {Powergraph: Distributed graph-parallel computation on natural
  graphs}.
\newblock In \emph{OSDI}, pages 17--30, 2012.

\bibitem[Hindman et~al.(2011)Hindman, Konwinski, Zaharia, Ghodsi, Joseph, Katz,
  Shenker, and Stoica]{hindman2011mesos}
B.~Hindman, A.~Konwinski, M.~Zaharia, A.~Ghodsi, A.~D. Joseph, R.~Katz,
  S.~Shenker, and I.~Stoica.
\newblock {Mesos: A Platform for Fine-grained Resource Sharing in the Data
  Center}.
\newblock In \emph{NSDI}, Berkeley, CA, USA, 2011.

\bibitem[Hwang and Wood(2013)]{hwang2013adaptive}
J.~Hwang and T.~Wood.
\newblock Adaptive performance-aware distributed memory caching.
\newblock In \emph{ICAC}, volume~13, pages 33--43, 2013.

\bibitem[Kalyvianaki et~al.(2016)Kalyvianaki, Fiscato, Salonidis, and
  Pietzuch]{kalyvianaki2016themis}
E.~Kalyvianaki, M.~Fiscato, T.~Salonidis, and P.~Pietzuch.
\newblock Themis: Fairness in federated stream processing under overload.
\newblock In \emph{SIGMOD}, pages 541--553. ACM, 2016.

\bibitem[Karger et~al.(1997)Karger, Lehman, Leighton, Panigrahy, Levine, and
  Lewin]{karger1997consistent}
D.~Karger, E.~Lehman, T.~Leighton, R.~Panigrahy, M.~Levine, and D.~Lewin.
\newblock Consistent hashing and random trees: Distributed caching protocols
  for relieving hot spots on the world wide web.
\newblock In \emph{STOC}, pages 654--663. ACM, 1997.

\bibitem[Katsipoulakis et~al.(2017)Katsipoulakis, Labrinidis, and
  Chrysanthis]{katsipoulakis2017holistic}
N.~R. Katsipoulakis, A.~Labrinidis, and P.~K. Chrysanthis.
\newblock A holistic view of stream partitioning costs.
\newblock \emph{VLDB}, 10\penalty0 (11), 2017.

\bibitem[Khayyat et~al.(2013)Khayyat, Awara, Alonazi, Jamjoom, Williams, and
  Kalnis]{khayyat2013mizan}
Z.~Khayyat, K.~Awara, A.~Alonazi, H.~Jamjoom, D.~Williams, and P.~Kalnis.
\newblock Mizan: a system for dynamic load balancing in large-scale graph
  processing.
\newblock In \emph{Proceedings of the 8th ACM European Conference on Computer
  Systems}, pages 169--182. ACM, 2013.

\bibitem[Koliousis et~al.(2016)Koliousis, Weidlich, Castro~Fernandez, Wolf,
  Costa, and Pietzuch]{koliousis2016saber}
A.~Koliousis, M.~Weidlich, R.~Castro~Fernandez, A.~L. Wolf, P.~Costa, and
  P.~Pietzuch.
\newblock Saber: Window-based hybrid stream processing for heterogeneous
  architectures.
\newblock In \emph{SIGMOD}, pages 555--569. ACM, 2016.

\bibitem[Lin et~al.(2009)]{lin2009curse}
J.~Lin et~al.
\newblock The curse of zipf and limits to parallelization: A look at the
  stragglers problem in mapreduce.
\newblock In \emph{7th Workshop on Large-Scale Distributed Systems for
  Information Retrieval}, volume~1, 2009.

\bibitem[Malewicz et~al.(2010)Malewicz, Austern, Bik, Dehnert, Horn, Leiser,
  and Czajkowski]{malewicz2010pregel}
G.~Malewicz, M.~H. Austern, A.~J. Bik, J.~C. Dehnert, I.~Horn, N.~Leiser, and
  G.~Czajkowski.
\newblock Pregel: a system for large-scale graph processing.
\newblock In \emph{SIGMOD}, pages 135--146. ACM, 2010.

\bibitem[Mirrokni et~al.(2016)Mirrokni, Thorup, and
  Zadimoghaddam]{mirrokni2016consistent}
V.~Mirrokni, M.~Thorup, and M.~Zadimoghaddam.
\newblock Consistent hashing with bounded loads.
\newblock \emph{arXiv preprint arXiv:1608.01350}, 2016.

\bibitem[Mitzenmacher et~al.(2001)Mitzenmacher, Sitaraman,
  et~al.]{mitzenmacher2001potc-survey}
M.~Mitzenmacher, R.~Sitaraman, et~al.
\newblock The power of two random choices: A survey of techniques and results.
\newblock In \emph{Handbook of Randomized Computing}, pages 255--312, 2001.

\bibitem[Nasir et~al.(2015{\natexlab{a}})Nasir, Morales, Garc\'ia-Soriano,
  Kourtellis, and Serafini]{nasir2015partial}
M.~A.~U. Nasir, G.~D.~F. Morales, D.~Garc\'ia-Soriano, N.~Kourtellis, and
  M.~Serafini.
\newblock Partial key grouping: Load-balanced partitioning of distributed
  streams.
\newblock \emph{arXiv preprint arXiv:1510.07623}, 2015{\natexlab{a}}.

\bibitem[Nasir et~al.(2015{\natexlab{b}})Nasir, Morales, Garc\'ia-Soriano,
  Kourtellis, and Serafini]{nasir2015power}
M.~A.~U. Nasir, G.~D.~F. Morales, D.~Garc\'ia-Soriano, N.~Kourtellis, and
  M.~Serafini.
\newblock The power of both choices: Practical load balancing for distributed
  stream processing engines.
\newblock In \emph{ICDE}, pages 137--148, April 2015{\natexlab{b}}.

\bibitem[Nasir et~al.(2016)Nasir, Morales, Kourtellis, and
  Serafini]{nasir2016two}
M.~A.~U. Nasir, G.~D.~F. Morales, N.~Kourtellis, and M.~Serafini.
\newblock When two choices are not enough: Balancing at scale in distributed
  stream processing.
\newblock In \emph{ICDE}, pages 589--600, May 2016.

\bibitem[Ousterhout et~al.(2013)Ousterhout, Wendell, Zaharia, and
  Stoica]{ousterhout2013sparrow}
K.~Ousterhout, P.~Wendell, M.~Zaharia, and I.~Stoica.
\newblock Sparrow: distributed, low latency scheduling.
\newblock In \emph{SOSP}, pages 69--84, 2013.

\bibitem[Park(2011)]{park2011multiplechoices}
G.~Park.
\newblock {A Generalization of Multiple Choice Balls-into-bins}.
\newblock In \emph{PODC}, pages 297--298, 2011.

\bibitem[Rahm and Marek(1995)]{rahm1995dynamic}
E.~Rahm and R.~Marek.
\newblock Dynamic multi-resource load balancing in parallel database systems.
\newblock In \emph{VLDB}, volume~95, pages 11--15. Citeseer, 1995.

\bibitem[Salihoglu and Widom(2013)]{salihoglu2013gps}
S.~Salihoglu and J.~Widom.
\newblock Gps: A graph processing system.
\newblock In \emph{Proceedings of the 25th International Conference on
  Scientific and Statistical Database Management}, page~22. ACM, 2013.

\bibitem[Schneider et~al.(2016)Schneider, Wolf, Hildrum, Khandekar, and
  Wu]{schneider2016dynamic}
S.~Schneider, J.~Wolf, K.~Hildrum, R.~Khandekar, and K.-L. Wu.
\newblock Dynamic load balancing for ordered data-parallel regions in
  distributed streaming systems.
\newblock In \emph{Middleware}, page~21. ACM, 2016.

\bibitem[Shah et~al.(2003)Shah, Hellerstein, Chandrasekaran, and
  Franklin]{shah2003flux}
M.~A. Shah, J.~M. Hellerstein, S.~Chandrasekaran, and M.~J. Franklin.
\newblock Flux: An adaptive partitioning operator for continuous query systems.
\newblock In \emph{ICDE}, pages 25--36. IEEE, 2003.

\bibitem[Suresh et~al.(2015)Suresh, Canini, Schmid, and Feldmann]{suresh2015c3}
L.~Suresh, M.~Canini, S.~Schmid, and A.~Feldmann.
\newblock C3: Cutting tail latency in cloud data stores via adaptive replica
  selection.
\newblock In \emph{NSDI}, pages 513--527, 2015.

\bibitem[Taft et~al.(2014)Taft, Mansour, Serafini, Duggan, Elmore, Aboulnaga,
  Pavlo, and Stonebraker]{2014marcoestore}
R.~Taft, E.~Mansour, M.~Serafini, J.~Duggan, A.~J. Elmore, A.~Aboulnaga,
  A.~Pavlo, and M.~Stonebraker.
\newblock E-store: Fine-grained elastic partitioning for distributed
  transaction processing systems.
\newblock \emph{VLDB}, 8\penalty0 (3):\penalty0 245--256, Nov. 2014.
\newblock ISSN 2150-8097.
\newblock \doi{10.14778/2735508.2735514}.
\newblock URL \url{http://dx.doi.org/10.14778/2735508.2735514}.

\bibitem[Vavilapalli et~al.(2013)Vavilapalli, Murthy, Douglas, Agarwal, Konar,
  Evans, Graves, Lowe, Shah, Seth, et~al.]{vavilapalli2013apache}
V.~K. Vavilapalli, A.~C. Murthy, C.~Douglas, S.~Agarwal, M.~Konar, R.~Evans,
  T.~Graves, J.~Lowe, H.~Shah, S.~Seth, et~al.
\newblock Apache hadoop yarn: Yet another resource negotiator.
\newblock In \emph{SCC}, page~5, 2013.

\bibitem[Xing et~al.(2005)Xing, Zdonik, and Hwang]{xing2005dynamic}
Y.~Xing, S.~Zdonik, and J.-H. Hwang.
\newblock Dynamic load distribution in the borealis stream processor.
\newblock In \emph{ICDE}, pages 791--802, 2005.

\bibitem[Yan et~al.(2015)Yan, Cheng, Lu, and Ng]{yan2015effective}
D.~Yan, J.~Cheng, Y.~Lu, and W.~Ng.
\newblock Effective techniques for message reduction and load balancing in
  distributed graph computation.
\newblock In \emph{WWW}, pages 1307--1317. ACM, 2015.

\bibitem[Zaharia et~al.(2008)Zaharia, Konwinski, Joseph, Katz, and
  Stoica]{zaharia2008improving}
M.~Zaharia, A.~Konwinski, A.~D. Joseph, R.~H. Katz, and I.~Stoica.
\newblock Improving mapreduce performance in heterogeneous environments.
\newblock In \emph{OSDI}, volume~8, page~7, 2008.

\end{thebibliography}

\end{document}